\documentclass[12pt,draftcls, onecolumn]{IEEEtran}

\usepackage{amssymb, graphicx, cite, amsmath,  psfrag, color, array, mathrsfs, setspace, algorithm, algpseudocode, mathtools, esint, multirow, epstopdf, chemfig,fixltx2e }
\newtheorem{lemma}{Lemma}
\usetikzlibrary{shapes, arrows, decorations.markings}

\usepackage{enumitem}
\usepackage{tikz}
\usepackage{pgfplots}

\usetikzlibrary{patterns} 

\allowdisplaybreaks

\makeatletter
\let\MYcaption\@makecaption
\makeatother
\usepackage[font=footnotesize]{subcaption}
\makeatletter
\let\@makecaption\MYcaption
\makeatother
\def\gammath{P_\text{th}}
\def\lsh{\psi} 
\def\dsd{d_\text{SD}}
\def\dsh{d_\text{SH}}
\def\dhd{d_\text{HD}}
\def\Prec{P_\text{r}}
\def\Pt{ P_\text{t}}
\def\Ps{\mathcal{P}^\text{Succ}_\text{Direct}}
\def\mylog{-K+10\,\alpha\,\log_{10}}

\def\sigmash{\sigma_\psi}
\def\Rcoop{R_\text{Coop}}
\def\rhd{R_\text{HD}}
\def\rsd{R_\text{SD}}
\def\rsh{R_\text{SH}}

\def\Pcoops{\mathcal{P}^\text{Succ}_\text{Coop}}
\def\PcoopsCi{\mathcal{P}^{\text{Succ},\mathcal{C},i}_\text{Coop}}
\def\PcoopsCa{\mathcal{P}^{\text{Succ},\mathcal{C},1}_\text{Coop}}
\def\PcoopsCb{\mathcal{P}^{\text{Succ},\mathcal{C},2}_\text{Coop}}
\def\PcoopsCc{\mathcal{P}^{\text{Succ},\mathcal{C},3}_\text{Coop}}

\def\psbar{\overline{\mathcal{P}}^\text{Succ}}

\def\psabar{\overline{\mathcal{P}}^\text{Succ}_\mathcal{A}}
\def\frk{\mathrm{f}_{r_k}(r)}

\def\psbbar{\overline{\mathcal{P}}^\text{Succ}_\mathcal{B}}

\def\lDai{\mathcal{L}_\text{Coop}^{\mathcal{D}_1,i}}
\def\uDai{\mathcal{U}_\text{Coop}^{\mathcal{D}_1,i}}
\def\lDbi{\mathcal{L}_\text{Coop}^{\mathcal{D}_2,i}}
\def\uDbi{\mathcal{U}_\text{Coop}^{\mathcal{D}_2,i}}

\begin{document}
\title{{ A New Approach for Helper Selection and Performance Analysis in Poisson CoopMAC Networks}}
\author{Homa Nikbakht, Amir Masoud Rabiei, and Vahid Shah-Mansouri,~\IEEEmembership{Member~IEEE} \\[1.5ex]
{\normalsize School of Electrical and Computer Engineering, College of Engineering \\
University of Tehran, P.O.~Box 14395--515, Tehran, Iran \\
E-mail: {\tt \fontsize{11}{12}\selectfont\{homanikbakht, rabiei, vmansouri\}@ut.ac.ir}\\[.25\baselineskip]}}

\maketitle

\begin{abstract}
The cooperative medium access control (CoopMAC) protocol in the presence of randomly-distributed nodes and shadowing is considered.  The nodes are assumed to be distributed according to a homogeneous two-dimensional Poisson point process.  A new approach is proposed for helper selection and throughput performance analysis which depends on the shadowing parameters as well as the distribution of helpers.  In the proposed protocol, the potential helpers are divided into several tiers based on their distances from the source and destination in a way that the lower the tier index, the higher its priority.  When there are several helpers of the same tier, the helper that is less affected by shadowing is chosen for cooperation.  Upper and lower bounds are derived for the average cooperative throughput of the proposed CoopMAC protocol.  It is observed that the proposed scheme readily outperforms the conventional CoopMAC protocol in having larger average throughput.  It is also seen that the cooperative throughput of the proposed scheme approaches the upper bound when the density of nodes increases.
\end{abstract}


\IEEEpeerreviewmaketitle

\section{Introduction}
The broadcast nature of the wireless medium is one of the most important features of wireless communication networks.
A direct consequence of this feature is that the transmission between any two nodes can be overheard by other nodes of the network.  
As a result, a source node can make use of the other nodes (referred to as \textit{helpers} in the sequel) to improve performance measures such as throughput, bit error rate (BER) and diversity gain.
\par In recent years, many studies have been conducted on cooperation between nodes in the medium access control (MAC) layer.
Some of the protocols proposed for the MAC layer are based on the IEEE 802.11b Standard \cite{IEEE802.11b}.  In this standard, a multirate scheme is employed for establishing a connection between two nodes in a wireless local area network (WLAN). 
An important drawback of the above multirate scheme is the fact that for transmitting the same amount of information, low-rate links make the channel busy for a longer time than the high-rate links.  A possible approach to overcome this deficiency is to make use of an appropriate helper to retransmit the source signal to destination and reduce the total transmission time, i.e., a cooperative MAC (CoopMAC) protocol is used \cite{Korakis}.
\par In a CoopMAC protocol, the nodes are assumed to be distributed in a given region and each node keeps a table (known as CoopTable) containing the information corresponding to the helpers that can possibly assist the source during its transmission \cite{Korakis}.  Before transmitting its packet, each node looks up the CoopTable to see if there is a helper that can improve the overall transmission rate.  If there are several such helpers, the one with the latest feedback time is chosen for cooperation. The feedback time is the latest time a successful transmission is observed from that helper.
In \cite{CODE}, a similar protocol to CoopMAC has been proposed in which two helpers with the latest feedback times are chosen to assist the source.  Interestingly, it is shown in reference \cite{forward} that a high-rate node assisting a low-rate node can improve its own throughput, delay and energy consumption.
This is because by forwarding a low-rate node's data, the high-rate node can gain access to a free channel in a shorter time to complete its own transmission \cite{forward}.  Observe that the creation and maintenance of the CoopTable needs additional memory at each node and increases the complexity of the system significantly \cite{CoopRTS/CTS}.  Moreover, the performance of CoopMAC protocols which are based on the CoopTable severely degrades when the helpers are mobile.  In order to address this problem, a new protocol based on CoopMAC has been proposed in \cite{WCNC} which separates the mobile helpers from the static ones by maintaining a history of the signal strength corresponding to each helper's overheard packets.

\par In \cite{PRCSMA}, a new cooperation protocol known as  persistent relay carrier sensing multiple access (PRCSMA) has been proposed which employs an automatic retransmission request (ARQ) scheme to enhance the overall performance of the IEEE 802.11 protocol.  In this protocol, each time a packet is received with error, the destination automatically transmits a claim for cooperation (CFC) packet to the other nodes, and requests for a retransmission of the original packet. In this protocol, all idle nodes can act as a potential helper as long as they satisfy a set of relay selection conditions.
PRCSMA protocol is known to improve the channel usage and to increase the transmission range \cite{PRCSMA}, however, it has poor bandwidth efficiency \cite{CoopRTS/CTS}.  
\par In \cite{distributed}, a variation of the CoopMAC protocol has been proposed in which cooperation takes place only when there exists a potential cooperative link which can provide a desired transmission rate that cannot be achieved by the direct link. This protocol, however, suffers from high computational complexity as the link capacity is estimated using instantaneous signal-to-noise power ratio (SNR), and the latter requires two channel state estimations (namely, source to helper and helper to destination channels) for each potential helper \cite{relayselection}.  Note that CoopMAC has been also investigated from other viewpoints in the literature.  In \cite{Wang15}, it is shown that the network lifetime can be improved by employing CoopMAC.  Also in \cite{Ju15}, a game theoretical approach has been proposed for analyzing a CoopMAC protocol with incentive design. 

\par An important assumption which seems to be less investigated in the literature is the effect of shadowing on the performance of the CoopMAC-based protocols  \cite{shadow}.  Shadowing occurs when there are obstacles that block the line-of-sight (LOS) path between two communicating nodes and can attenuate the transmitted signal power drastically.  Therefore, it is a major impairment in wireless networks and must be taken into account in the design and evaluation of these networks \cite{shadow,shadow2,shadownc}. In \cite{shadow}, a new CoopMAC protocol has been proposed and the effect of uncorrelated shadowing on the average number of nodes that can receive a packet with desired quality of service (QoS) has been examined.  In addition, the effect of correlated shadowing on the number of helpers that are capable of cooperation in a two-way network-coded (NC) relaying system has been studied in \cite{shadownc}.  
\par Motivated by the above facts, in this paper we propose a new helper selection scheme for a CoopMAC network whose nodes are distributed according to a homogeneous two-dimensional Poisson point process (PPP) with a fixed density \cite{Wang_2011,behnad_TCOM_2013}.  To the authors knowledge, the effect of random spatial distribution of the nodes on the overall throughput of the CoopMAC networks seems to have received little attention in the literature.   We assume the communication between any two nodes is subject to path loss and shadowing and derive exact expression for the throughput of the direct link between two arbitrary nodes in a random CoopMAC network.\footnote{We assume in the following that in a \textit{Poisson CoopMAC network}, the locations of the nodes constitute a two-dimensional PPP with a fixed density.}  We also find upper and lower bounds for the throughput of a cooperative link making use of our proposed helper selection scheme in the presence of shadowing. Our numerical results demonstrate that the proposed CoopMAC has superior throughput performance, and its throughput is only slightly smaller than the upper bound in all the examined scenarios.
\par The rest of this paper is organized as follows. In Section \ref{system_model}, the system model is introduced. We classify the links based on their corresponding source-destination distance and propose our helper selection scheme and data transmission framework in Section \ref{helper selection}.  In Sections \ref{analytical}, \ref{boundsC} and \ref{boundsD}, a theoretical throughput analysis is presented for each of the link types introduced in Section \ref{helper selection}.   Numerical results are provided in Section \ref{simulation} to verify the superiority of our proposed scheme over the conventional CoopMAC protocol.  Concluding remarks are presented in Section \ref{conclude}.

\begin{figure}[!t]
\vspace*{-7ex}
\centering
\begin{subfigure}[t]{.5\linewidth}
  \centering
    \hspace*{32pt}
 \begin{tikzpicture}[scale=2.5, >=stealth]
\centering
\tikzstyle{every node}=[draw,shape=circle, node distance=1cm];
\node (s0) at (0,0 ) {S};
\node (d0) at (2,0) {D};
\node (h0) at (1,1 ) {H};
\draw   [thick,->] (s0)-- node [draw=none,  text width=3.0cm, shape=rectangle, pos=0.5, yshift=0.2cm,xshift=0.1cm ,midway, fill=none, node distance=1cm] {\small\text{CoopRTS}}  (d0);
\draw   [thick,->] (s0)-- node [draw=none, rotate = 45,  text width=3.0cm, shape=rectangle, pos=0.5, yshift=-0.25cm,xshift=0.6cm ,midway, fill=none, node distance=1cm] {\small\text{CoopRTS}}  (h0);
\draw   [thick,->] (1.85,-0.08)-- node [draw=none, rotate = 0,  text width=3.0cm, shape=rectangle, pos=0.5, yshift=-0.2cm,xshift=2.6cm ,midway, fill=none, node distance=1cm] {\small\text{CTS}}  (0.15,-0.08);
\draw   [thick,->] (0.85,0.95)-- node [draw=none, rotate = 45,  text width=3.0cm, shape=rectangle, pos=0.5, yshift=0.2cm,xshift=2.2cm ,midway, fill=none, node distance=1cm] {\small\text{HTS}}  (0.05,0.15);
\draw   [thick,->] (1.15,0.95)-- node [draw=none, rotate = -45,  text width=3.0cm, shape=rectangle, pos=0.5, yshift=0.2cm,xshift=0.2cm ,midway, fill=none, node distance=1cm] {\small\text{HTS}}  (1.95,0.144);
\end{tikzpicture}
  \caption{}
  \label{fig1-1}
\end{subfigure}~~%
\begin{subfigure}[t]{.5\linewidth}
  \centering
  \hspace*{25pt}
\begin{tikzpicture}[scale=2.5, >=stealth]
\centering
\tikzstyle{every node}=[draw,shape=circle, node distance=1cm];
\node (s0) at (0,0 ) {S};
\node (d0) at (2,0) {D};
\node (h0) at (1,1 ) {H};
\draw   [thick,->] (d0)-- node [draw=none,  text width=3.0cm, shape=rectangle, pos=0.5, yshift=-0.375cm,xshift=2.5cm ,midway, fill=none, node distance=1cm] {\small\text{ACK}\vspace*{2pt}}  (s0);
\draw   [thick,->] (s0)-- node [draw=none, rotate = 45,  text width=3.0cm, shape=rectangle, pos=0.5, yshift=-0.25cm,xshift=0.6cm ,midway, fill=none, node distance=1cm] {\small\text{Data}}  (h0);
\draw   [thick,->] (h0)-- node [draw=none, rotate = -45,  text width=3.0cm, shape=rectangle, pos=0.5, yshift=0.2cm,xshift=0.2cm ,midway, fill=none, node distance=1cm] {\small\text{Data}}  (d0);
\end{tikzpicture}
  \caption{}
  \label{fig1-2}
\end{subfigure}
\caption{Illustration of (a) control frame exchange, and (b)~data frame exchange, in a typical CoopMAC link.}
\label{figpr1}
\end{figure}
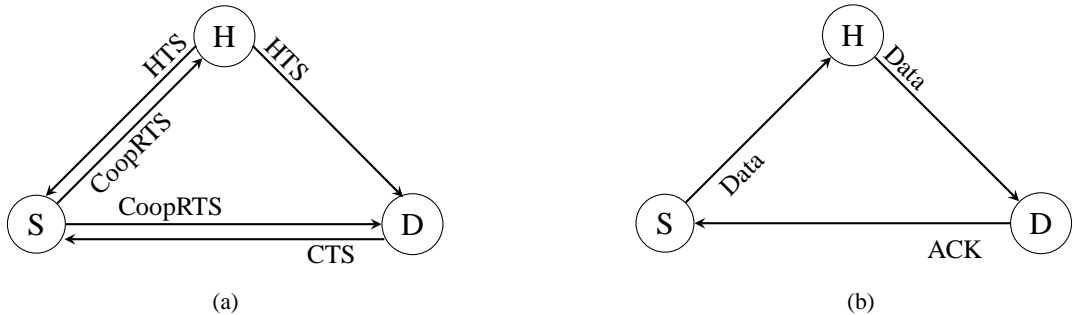
\section{System Model}
\label{system_model}
In this section, we first provide a brief description on how a CoopMAC protocol works.  Then, the effect of shadowing on the probability of a successful transmission is examined.
\subsection{The CoopMAC Protocol} \label{coopmac}
We consider a wireless network whose nodes constitute a homogeneous two dimensional PPP with density $\lambda$.  The nodes communicate with each other according to a carrier sense multiple access with collision avoidance (CSMA/CA) scheme in the IEEE 802.11b Standard in distributed coordination function (DCF) mode \cite{IEEE802.11b}. We assume the propagation delay is small enough to ensure that CSMA/CA provides a collision-free environment for destination. 
Our network makes use of the RTS/CTS technique, in which each node can distinguish whether the received packet is for itself or should be forwarded to another node. 

A typical CoopMAC link making use of an RTS/CTS scheme is shown in Fig. \ref{figpr1} \cite{Korakis} . Before any communication, the best potential helper must be selected based on a set of criteria.  Then, the source node transmits a cooperative RTS (CoopRTS) packet to the best helper and reserves a channel.  If the helper is willing to cooperate, it sends back a helper-ready-to-send (HTS) packet. Moreover, the destination node confirms the reservation of a channel by transmitting a CTS packet to the source \cite{Korakis}. If the source node receives both HTS and CTS packets, the cooperation starts and the helper forwards the data packets to the destination through cooperative link.  If only the CTS is received, the transmission is made through the direct link.  If neither CTS nor HTS packets are received during a specific time, a timeout occurs and the transmission is declared as failed. 
The transmission is considered as successful when the source node receives an acknowledgment (ACK) packet from destination.
\par We assume in the sequel that $\dsh$, $\dhd$ and $\dsd$ represent the source-to-helper, helper-to-destination and source-to-destination distances, respectively.\footnote{Throughout this paper, all distances are expressed in meters.} Moreover, $\rsh$, $\rhd$ and $\rsd$ denote the transmission rates of source-helper (S--H), helper-destination (H--D) and source-destination (S--D) links, respectively.
\begin{table}[!t]
\caption{The link types and their corresponding transmission rates in the \textrm{IEEE 802.11b} Standard (BER $\ge 10^{-5}$).}
\centering
    \begin{tabular}{ | c | c | c |}
    \hline
    Link Type &$\dsd$ (meter) & Transmission Rate \\ \hline
    $\mathcal{A}$ &$0 \le \dsd < 48.2$ &11 Mbps  \\ \hline
    $\mathcal{B}$ & $48.2 \le \dsd < 67.1$ & 5.5 Mbps \\ \hline
    $\mathcal{C}$ &$67.1 \le \dsd < 74.7$  & 2 Mbps \\ \hline
    $\mathcal{D}$ &$74.7 \le \dsd \le 100$  &1 Mbps \\
    \hline
    \end{tabular}
\label{T1}
\end{table}
\subsection{Effect of Shadowing on Successful Transmission Probability}
\par Shadowing is referred to case where the received signal power is affected by the objects obstructing the path between transmitter and receiver. In order to model the path loss plus shadowing effects, we assume that the received power in dB at destination node is given by \cite[eq. 2.51]{Goldsmith}
\begin{equation}\label{Preceived}
\Prec = \Pt  +K-10\,\alpha\,\log_{10} (\dsd)+\lsh,
\end{equation}
where $\Pt$ is the transmitted power in dB and is assumed to be the same for all nodes (including the helpers), $K$ is a constant in dB which depends on the antenna characteristics, $\alpha$ is the path loss exponent usually between $2$ and $7$, 
and $\lsh$ is a Gaussian random variable in dB units which represents the effect of shadowing and has mean zero and standard deviation $\sigmash$.
Depending on the quality of service (QoS) requirements, a threshold for the received power in dB, i.e., $\gammath$, is defined at the destination.  Therefore, the probability of a successful transmission  through the direct link between two nodes at distance $\dsd$ equals 
\begin{equation}\label{Psucc}
\Ps=\Pr\{ \Prec \ge \gammath\}.
\end{equation}
Substituting for $\Prec$ from \eqref{Preceived} into \eqref{Psucc}, we obtain
\begin{subequations}
\begin{align} \notag
{\Ps} &= \Pr\{\lsh \ge \gammath-\Pt \mylog(\dsd)\} \\ \label{Psucc2q}
	        &=\mathbb{Q}\left(\nu+\mu \log_{10}(\dsd)\right)
\shortintertext{where}
		\nu & = \frac{\gammath-\Pt-K}{\sigmash} \\
		\mu & = \frac{10\alpha}{\sigmash}
\end{align}
\end{subequations}
and
$\mathbb{Q}\left( x \right)  \triangleq  \frac{1}{{\sqrt {2\pi } }}\int_x^\infty  {{e^{ - {u^2}/2}}du}$
is the Gaussian $\mathbb{Q}$--function.  Note that in our treatment, the maximum effective transmission range equals $100$ meters. Therefore, it is reasonable to assume that the argument of $\mathbb{Q}$--function in \eqref{Psucc2q} is always negative, or analogously $\Ps> 0.5$ for $0<\dsd\le 100$.
\par Similarly, for a cooperative link (i.e., a link whose source and destination nodes communicate through a helper) the probability of a successful transmission can be obtained as
\begin{subequations}
\begin{align}  \label{PSuccessDef}
\Pcoops & = \Pr\{P_{\text{r,\,H}} \ge P_\text{th}\}\; \Pr\{P_{\text{r,\,D}}\ge P_\text{th}\}=\mathbb{G}(\dsh,\dhd) \\
\shortintertext{where} \label{PSuccessDefb}
\mathbb{G}(\ell_1,\ell_2) & = \mathbb{Q}\left(\nu+\mu \log_{10}(\ell_1)\right) \mathbb{Q}\left(\nu+\mu \log_{10}(\ell_2)\right)
\end{align}
\end{subequations}
and $P_{\text{r,\,H}}$ and $P_{\text{r,\,D}}$ are the received powers at the helper and destination, respectively.

\begin{table}[!t]
	\caption{The helper tiers and transmission rates for a Type $\mathcal{C}$ link.}
	\centering
	\renewcommand{\arraystretch}{1.25}
    \tabcolsep=0.1cm
	\label{table2}
	\begin{tabular}{|c|c|c|c|c|c|}\hline
		Tier& \; $\dsh$ (meter)&\; $\rsh$ (Mbps)&\; $\dhd$ (meter)&\; $\rhd$ (Mbps)&$\Rcoop$ (Mbps)\\\hline
		1&\;$[0,48.2)$&\;$11$&\;$[0,48.2)$&\;$11$&\;$5.5$\\\hline
		\multirow{2}{*}{2}&\;$[0,48.2)$&\;$11$&\;$[48.2,67.1)$&\;$5.5$&\;\multirow{2}{*}{$3.67$}\\\cline{2-5}
		&\;$[48.2,67.1)$&\;$5.5$&\;$[0,48.2)$&\;$11$&\\\hline
		3&\;$[48.2,67.1)$&\;$5.5$&\;$[48.2,67.1)$&\;$5.5$&\;$2.75$\\\hline
	\end{tabular}	
	\label{T2}	
\end{table}

\section{Network Classification and Best Helper Selection Algorithm}
\label{helper selection}
In the IEEE 802.11b Standard, the closer the source and destination nodes the higher the transmission rate \cite{IEEE802.11b}.  As a result, in a CoopMAC network based on the IEEE 802.11b Standard, we can define four types of links depending on the distance between the source and destination nodes as illustrated in Table \ref{T1}. In what follows we investigate whether a helper can increase the transmission rate corresponding to each of the link types listed in Table \ref{T1} or not.

\subsubsection{Type $\mathcal{A}$ Links}
The transmission rate for this type is $11$ Mbps which is the maximum transmission rate in the IEEE 802.11b Standard.  In consequence, a helper is not used for the links of this type.

\subsubsection{Type $\mathcal{B}$ Links}
Similar to Type $\mathcal{A}$ links, for Type $\mathcal{B}$ links a helper cannot improve the transmission rate.  To show this, we assume that the source wants to transmit $L$ bits of information to destination through a helper.  Denoting the transmission times of S--H and H--D links by $t_\text{SH}$ and $t_\text{HD}$, respectively, we can obtain the cooperation time as
\begin{equation}\label{Tcoop}
{t_\text{Coop}}=t_\text{SH}+t_\text{HD}.
\end{equation}
It is clear that the time taken for transmission of a sequence of $L$ bits\footnote{This sequence is assumed to include the RTS and CTS packets as well.} over a link with a transmission rate of $R$ bps equals $L/R$.  Thus, the overall rate of the cooperative link (i.e., S--H--D link), is given by \cite[eq. (1)]{WCNC}
\begin{equation}\label{RateCoop}
\Rcoop =\frac{L}{t_\text{Coop}}= \frac{L}{{\frac{L}{{{\rsh}}} + \frac{L}{{{\rhd}}}}} =
\frac{\rsh\,\rhd}{\rsh+\rhd}
\end{equation}
where $\rsh$ and $\rhd$ are the S--H and H--D link rates.  In order to obtain the maximum value of $\Rcoop$, one should minimize the denominator of the fraction on the right of \eqref{RateCoop}.  Clearly, the minimum is attained when $\rsh=\rhd=11$ Mbps and, thus, the maximum rate of the cooperative link becomes $5.5$ Mbps which equals the transmission rate of the direct link.  Hence, even in the best-case scenario a helper cannot improve the transmission rate of a Type $\mathcal{B}$ link.
%
%
\begin{table}[!t]
	\caption{The helper tiers and transmission rates for a Type $\mathcal{D}$ link.}
	\centering
	\renewcommand{\arraystretch}{1.25}
    \tabcolsep=0.1cm
	\label{table2}
	\begin{tabular}{|c|c|c|c|c|c|}\hline
		Tier& \; $\dsh$ (meter)&\; $\rsh$ (Mbps)&\; $\dhd$ (meter)&\; $\rhd$ (Mbps)&$\Rcoop$ (Mbps)\\\hline
		1&\;$[0,48.2)$&\;$11$&\;$[0,48.2)$&\;$11$&\;$5.5$\\\hline
		\multirow{2}{*}{2}&\;$[0,48.2)$&\;$11$&\;$[48.2,67.1)$&\;$5.5$&\;\multirow{2}{*}{$3.67$}\\\cline{2-5}
		&\;$[48.2,67.1)$&\;$5.5$&\;$[0,48.2)$&\;$11$&\\\hline
		3&\;$[48.2,67.1)$&\;$5.5$&\;$[48.2,67.1)$&\;$5.5$&\;$2.75$\\\hline
		\multirow{2}{*}{4}&\;$[0,48.2)$&\;$11$&\;$[67.1,74.7)$&\;$2$&\;\multirow{2}{*}{$1.69$}\\\cline{2-5}
		&\;$[67.1,74.7)$&\;$2$&\;$[0,48.2)$&\;$11$&\\\hline
		\multirow{2}{*}{5}&\;$[48.2,67.1)$&\;$5.5$&\;$[67.1,74.7)$&\;$2$&\;\multirow{2}{*}{$1.47$}\\\cline{2-5}
		&\;$[67.1,74.7)$&\;$2$&\;$[48.2,67.1)$&\;$5.5$&\\\hline
	\end{tabular}	
	\label{T3}	
\end{table}

\subsubsection{Type $\mathcal{C}$ Links}
For the links of this type, there are four different cases where cooperation is beneficial, i.e., the overall transmission rate is greater than $2$ Mbps.  These cases lead 
to three different helper tiers, viz.,
\setlist[description]{font=\normalfont\itshape}
\begin{description}[leftmargin=40pt]
\item [Tier 1:] Both $\dsh$ and $\dhd$ are less than 48.2, and $\Rcoop$ is equal to $5.5$ Mbps.
\item [Tier 2:] Either $\dsh$ or $\dhd$ is less than 48.2 and the other is in $[48.2,67.1)$ range.  $\Rcoop$ equals $3.67$ Mbps.
\item [Tier 3:] Both $\dsh$ and $\dhd$ are in $[48.2,67.1)$ range, and $\Rcoop=2.75$ Mbps.
\end{description} 
The above definitions are summarized in Table \ref{T2}.  
Clearly, for Type $\mathcal{C}$ links, a helper is useful only when neither $\rsh$ nor $\rhd$ are smaller than $5.5$ Mbps. Note that the larger the tier index the smaller the cooperative rate ($\Rcoop$).


\subsubsection{Type $\mathcal{D}$ Links} \label{sysmodD}
There are eight cases where a helper can improve the overall transmission rate of a Type $\mathcal{D}$ link.  As shown in Table \ref{T3}, these cases result in five different tiers of helpers. Observe that the specifications of Tiers $1$ through $3$ helpers are the same for both Type $\mathcal{C}$ and Type $\mathcal{D}$ links. The specifications of Tier $4$ and Tier $5$ helpers are given in Table \ref{T3}.
Again, as the tier index increases, the corresponding cooperative rate decreases.  Note, importantly, that a Type $\mathcal{D}$ link is established between two nodes whose distance is between $74.7$ and $100$ meters.  Thus, a Tier 1 helper does not exist when the source and destination nodes are more than $96.4$ meters apart.

In summary, a helper in a CoopMAC network is beneficial when the following two conditions are satisfied:
\begin{enumerate}
  \item The link between the source and destination nodes is of Type $\mathcal{C}$ or $\mathcal{D}$.
  \item The transmission rate of the cooperative link ($\Rcoop$) exceeds that of the direct link.
\end{enumerate}
\par Suppose now that the S--D link is of Type $\mathcal{C}$.  Then, a list of Tier 1 helpers denoted by $\mathcal{H}_1$, is created. If $\mathcal{H}_1\ne\varnothing$, the helper with the largest $\Pcoops$ in $\mathcal{H}_1$, is selected for cooperation.  The CoopRTS/CTS handshake is then accomplished as explained in Subsection \ref{coopmac}.  When $\mathcal{H}_1$ is empty, a list of Tier 2 helpers (i.e., $\mathcal{H}_2$) is formed and, again, the helper with the largest $\Pcoops$ in $\mathcal{H}_2$, is selected for cooperation.  Similarly, when both $\mathcal{H}_1$ and $\mathcal{H}_2$ are empty, a list of Tier 3 helpers is created, and cooperation is done through a helper in $\mathcal{H}_3$ that has the largest $\Pcoops$. A detailed explanation of the helper selection and transmission procedures for a Type $\mathcal{C}$ link is given in Algorithm \ref{alg1}.  Note that Algorithm \ref{alg1} can be readily modified to be used for the case where the S--D link is of Type $\mathcal{D}$.
\begin{algorithm}[t]
\caption{Best helper selection and data transmission for a Type $\mathcal{C}$ link}\label{best helper1}
\begin{algorithmic}[1]
\State \textbf {Initialization:}\label{op1}
\If {\text{there is any Tier 1 helper}} \label{op2}
\State \text{create $\mathcal{H}_1$, i.e., \textit{a list of Tier 1 helpers}} \label{op3}
\EndIf
\State \text{\par select the helper from $\mathcal{H}_1$ with the largest $\mathbb{G}(\dsh,\dhd)$ and remove it from $\mathcal{H}_1$} \label{op4}
 \State \text{send a  {CoopRTS} packet to the helper chosen in Step \ref{op4}} \label{op6}
 \If{ \text{an {HTS} packet is received}}  \textbf{goto} Step \ref{op21}
\ElsIf{ \text{$\mathcal{H}_1$} is not empty} \textbf {goto} Step \ref{op4}
 \EndIf
\If { \text{there is any Tier 2 helper}}\label{op8}
\State \text{create $\mathcal{H}_2$, i.e., \textit{a list of Tier 2 helpers}} \label{op9}
\EndIf
\State \text{select the helper from $\mathcal{H}_2$ with the largest $\mathbb{G}(\dsh,\dhd)$ and remove it from $\mathcal{H}_2$} \label{op10}
 \State \text{send a {CoopRTS} packet to the helper chosen in Step \ref{op10}} \label{op12}
 \If{ \text{an {HTS} packet is received}}  \textbf{goto} Step \ref{op21} 
 \ElsIf{ \text{$\mathcal{H}_2$} is not empty} \textbf{goto} Step \ref{op10}
 \EndIf
\If { \text{there is any Tier 3 helper}}\label{op14}
\State \text{create $\mathcal{H}_3$, i.e., \textit{a list of Tier 3 helpers}} \label{op15}
\EndIf
\State \text{ select the helper from $\mathcal{H}_3$ with the largest $\mathbb{G}(\dsh,\dhd)$ and remove it from $\mathcal{H}_3$} \label{op16}
\State \text{send a CoopRTS packet to the helper chosen in Step \ref{op16}} \label{op18}
 \If{ \text{an {HTS} packet is received}}  \textbf{goto} Step \ref{op21}
 \ElsIf{ \text{$\mathcal{H}_3$} is not empty} \textbf {goto} Step \ref{op16}
 \EndIf
\State \text{send an {RTS} packet} \label{op20}
\If{ \text{a {CTS} packet is not received}}   \label{op21}
\State \text{perform a random backoff and \textrm{\bf{goto}} Step \ref{op1}} \label{op22}
\EndIf
\State \text{send data} \label{op23}
\If{ \text{an {ACK} packet is not received}} 
\State \text{perform a random backoff and \textrm{\bf{goto}} Step \ref{op1}} \label{op24}
\EndIf
\State  \textbf{Transmission Complete} \label{op25}
\end{algorithmic}
\label{alg1}
\end{algorithm}

\section{Performance Analysis in the Presence of Shadowing} \label{analytical}
In this section, we evaluate the throughput of the link types shown in Table \ref{T1} in the presence of shadowing and path loss.  For a link whose transmission rate equals $R$ bps the throughput is given by
\begin{equation}\label{throughput}
\mathcal{T}  = \psbar\cdot R
\end{equation}
where $\psbar$ is the average probability of a successful transmission through this link.
We use eq. \eqref{throughput}  in the following to find the throughputs of the link types presented in Table \ref{T1}.
\subsection{Type $\mathcal{A}$ and Type $\mathcal{B}$ Links}
As mentioned earlier, for Type $\mathcal{A}$ and Type $\mathcal{B}$ links a helper cannot improve the overall transmission rate.  Therefore, for these link types a helper is not used.  Assume that the source node is the $k$th nearest neighbor of the destination node and $\dsd=r_k$. When the nodes are distributed according to a two dimensional homogeneous PPP with density $\lambda$, the probability density function (PDF) of $r_k$ is given by \cite{Haenggi2005}
\begin{equation}\label{frk1}
\frk = 2e^{-\lambda \pi r^2}\frac{\big(\lambda \pi r^2\big)^k}{r(k-1)!} \mathrm{u}(r)
\end{equation}
where $\mathrm{u}(r)$ is the unit step function.
Thus, using \eqref{Psucc2q} along with \eqref{frk1}, we can find the average probability of a successful transmission for Type $\mathcal{A}$ and Type $\mathcal{B}$ links as $\psabar=\mathbb{H}(0,48.2)$ and $\psbbar=\mathbb{H}(48.2,67.1)$, respectively, where
\begin{equation}
\mathbb{H}(r_\text{min},r_\text{max}) \triangleq \int_{r_\text{min}}^{r_\text{max}} {\mathbb{Q}\left(\nu+\mu \log_{10}(r)\right) \frk \mathrm{d}r}.
\end{equation}
Hence, using Table \ref{T1} and eq. \eqref{throughput} we can obtain the throughputs of Type $\mathcal{A}$ and Type $\mathcal{B}$ links as
\begin{align}\label{throughputA}
\mathcal{T}^\mathcal{A} &= {\psabar}\times 11~(\text{Mbps}) \\ \label{throughputB}
\mathcal{T}^\mathcal{B} &= {\psbbar}\times 5.5~(\text{Mbps})
\end{align}
respectively.
\subsection{Type $\mathcal{C}$  and Type $\mathcal{D}$ Links} \label{SecTypeC}
For Type $\mathcal{C}$ and $\mathcal{D}$ links a helper may or may not be used as explained in Section \ref{system_model}.  Assume now that there is no helper for cooperation and the transmission is made through the direct link.  Then, similar to Type $\mathcal{A}$ and $\mathcal{B}$ links, the throughput of Type $\mathcal{C}$ and $\mathcal{D}$ links can be obtained, respectively, as
\begin{align}\label{throughputC}
\mathcal{T}^\mathcal{C}_\text{Direct} &= \mathbb{H}(67.1,74.7) \times 2~(\text{Mbps}) \\
\mathcal{T}^\mathcal{D}_\text{Direct} &= \mathbb{H}(74.7,100) \times 1~(\text{Mbps}).
\end{align}
When a helper is available for cooperation, the resulting throughput (referred to as cooperative throughput) equals $\Rcoop\,\Pcoops$ where $\Rcoop$ is given in Table \ref{T2} for a Type $\mathcal{C}$ link and in Table \ref{T3} for a Type $\mathcal{D}$ link, and $\Pcoops$ was defined in \eqref{PSuccessDef}.  Observe that the cooperative throughput depends on the link type as well as the tier of the helper.
Since the helpers are randomly distributed in the S--D plane, the average cooperative throughput is obtained by averaging $\Rcoop\,\Pcoops$ over the spatial distribution of the helper nodes.  This can become quite complicated as it requires the joint PDF of $\dsh$ and $\dhd$ which is not easy to obtain.  Moreover, the final result involves a three-fold integration which is difficult to evaluate.  Therefore, in the next two sections we derive upper and lower bounds on $\Pcoops$ for Type $\mathcal{C}$ and $\mathcal{D}$ links and use these bounds to subsequently derive upper and lower bounds on the average cooperative throughputs of these links.
\par Before proceeding further, we establish a fact which will be used in the sequel to obtain the probability that a helper of a given tier can be found for a  Type $\mathcal{C}$ or $\mathcal{D}$ link.  Assume that we have a field of nodes distributed in a region $\mathscr{R}$ according to a two-dimensional PPP with density $\lambda$.  Also assume that $\mathscr{A}$ is a subregion of $\mathscr{R}$, i.e., $\mathscr{A} \subseteq \mathscr{R}$. Then the probability that a node X is located in $\mathscr{A}$, provided that it is located in $\mathscr{R}$ equals  \cite[Def. 3.2--(ii)]{Moller}
\begin{equation} \label{s0}
\Pr\{\text{X} \in \mathscr{A}|\text{X} \in \mathscr{R}\} = \frac{\mathcal{S}(\mathscr{A})}{\mathcal{S}(\mathscr{R})}
\end{equation}
where $\mathcal{S}(\mathscr{A})$ and $\mathcal{S}(\mathscr{R})$ are the surface areas of $\mathscr{A}$ and $\mathscr{R}$, respectively.
%
\section{Bounds on The Average Throughput of a Type $\mathcal{C}$ Link} \label{boundsC}
In this section we evaluate the maximum and minimum cooperative throughputs of a Type $\mathcal{C}$ link.  In the following subsections, we assume $\PcoopsCi$ and $\mathcal{T}_\text{Coop}^{\mathcal{C},i}$ to be, respectively, the probability of a successful transmission and the throughput of a Type $\mathcal{C}$ link making use of a Tier $i$ helper where $i=1,\,2 \text{ and } 3$ as given in Table \ref{T2}.
\subsection{The Cooperative Throughput Using a Tier 1 Helper}  \label{SecTier1}
\begin{figure}[!t]
\centering
\begin{tikzpicture}[>=stealth, scale=0.85, every node/.style={transform shape}]
\tikzstyle{every node}=[draw,shape=circle, node distance=1cm];
\draw [fill =black](-1.75, 0) circle (0.1);
\draw [fill =black](1.75, 0) circle (0.1);
\node [draw=none](s) at (-1.95,0.3 ) {S};
\node [draw=none](d) at (1.75,0.3 ) {D};

\begin{scope}
\clip (-1.75, 0) circle (2.5);
\clip (1.75, 0) circle (2.5);
\fill[fill=none] (-4,2.5)
rectangle (4,-2.5);
\pattern [pattern=north east lines,pattern color=gray!40]
 (-4,2.5)
rectangle (4,-2.5);
\end{scope}
\draw (-1.75, 0) circle (2.5);
\draw [dashed,thick] (1.75,0) circle (3.5);
\draw (1.75, 0) circle (2.5);
\node [draw=none] (v1) at (-0.35,0.8) {$\mathscr{U}_1$};
\node [draw=none] (v1) at (4.7,2.4) {$\mathscr{W}$};
\draw [fill =black](0, 1.81) circle (0.07);
\draw [fill =black](0, -1.82) circle (0.07);
\draw [fill =black](0.43,0.67) circle (0.07);
\draw [fill =black](0.43,0) circle (0.07);
\draw (0,1.8)--(0,-1.8);
\node [draw=none] (v1) at (0.3,0.8) {\small${\text{H}^{*}}$};
\node [draw=none] (v1) at (0.5,-0.2) {\small {H$^\perp$}};
\node [draw=none] (v1) at (0,2.1) {\small$\text{K}_1$};
\node [draw=none] (v1) at (0,-2.15) {\small${\text{K}_2}$};
\draw [->] (1.75,0) -- node [draw=none,  text width=1cm, pos=0.5, yshift=-0.5cm ,above, fill=none, node distance=2cm] {\small$48.2$\,m} (4.25,0);
\draw [->] (-1.75,0) -- node [draw=none,  text width=1cm, pos=0.5, yshift=-0.5cm ,above, fill=none, node distance=1cm] {\small $48.2$\,m} (-4.25,0);
\draw [->] (1.75,0) -- node [draw=none,  text width=0.5cm, pos=0.5, yshift=-0.25cm ,midway, fill=none, node distance=1cm] {$r_k$} (4.6,-2);
\draw (-1.75,0)--(1.75,0);
\draw [dashed](0.43,0.67)--(0.43,0);
\draw [->] (0.43,0.67) -- (1.75,0);
\draw [->] (0.43,0.67) -- (-1.75,0);
\end{tikzpicture}
\caption{A typical Type $\mathcal{C}$ link making use of a Tier 1 helper, $67.1\le r_k\le74.7$.}
\label{fig2}
\end{figure}
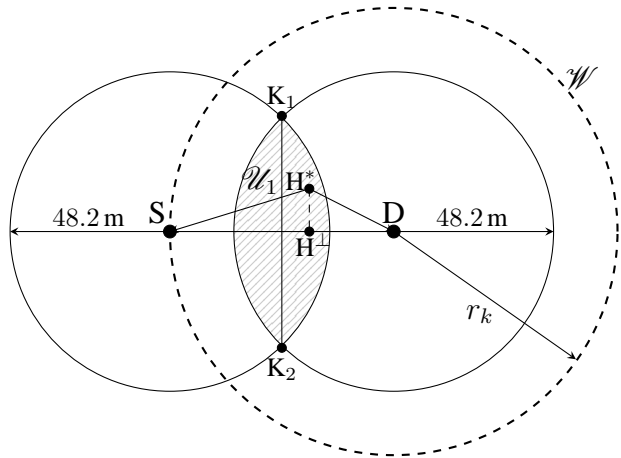
Fig. \ref{fig2} illustrates a Type $\mathcal{C}$ link in which the source and destination nodes communicate through a Tier 1 helper as shown in Table \ref{T2}.  Clearly, the helper has to be located in the shaded area, $\mathscr{U}_1$, i.e., the intersection of two circles centered at S and D both with radius $48.2$.  Since S is the $k$th nearest neighbor of D, there should be exactly $k-1$ nodes in a circle centered at D with radius $r_k$ (this region is referred to as $\mathscr{W}$ in Fig. \ref{fig2}).  Denoting by $\mathbb{N}(\mathscr{U}_1)$ the number of nodes in $\mathscr{U}_1$ and using the fact that $\mathscr{U}_1 \subseteq \mathscr{W}$ along with \eqref{s0}, one can find the probability that at least one Tier 1 helper is available for a Type $\mathcal{C}$ link as
\begin{align}
\mathcal{P}_{\mathcal{C},1}=\Pr\{\mathbb{N}(\mathscr{U}_1)\ge1\} &= 1-\Pr\{\mathbb{N}(\mathscr{U}_1)=0\} \notag \\
& = 1-\Pr\{\mathbb{N}(\mathscr{W}\!-\!\mathscr{U}_1)=k-1\} \notag \\\label{PC,1}
& = 1-\bigg(\frac{\mathcal{S}(\mathscr{W})\!-\!\mathcal{S}(\mathscr{U}_1)}{\mathcal{S}(\mathscr{W})}\bigg)^{k-1}
 = 1-\bigg(1-\frac{\mathcal{S}(\mathscr{U}_1)}{\pi\,\!r_k^2}\bigg)^{\!k-1}
\end{align}
where $\mathcal{S}(\mathscr{U}_1) = \mathbb{A}(48.2,48.2,r_k)$ and $\mathbb{A}(r_1,r_2,\ell)$ is the surface area of the intersection of two circles with radii $r_1$ and $r_2$ whose centers are $\ell$ meters apart, that is \cite{Korakis}
\begin{subequations}
\begin{equation}\label{overlap}
\mathbb{A}(r_1,r_2,\ell) = r_{1}^{2}\, \arcsin\Big(\frac{h}{r_1}\Big) +r_{2}^{2}\, \arcsin\Big(\frac{h}{r_2}\Big) - h\ell,
\end{equation}
where
\begin{equation}\label{overlaph}
h = \frac{\sqrt{2r_{1}^{2}r_{2}^{2} + 2(r_{1}^{2} +r_{2}^{2})\ell^{2}-(r_{1}^{4} +r_{2}^{4}) \ell^{4}}}{2\ell}.
\end{equation}
\end{subequations}
We now state and prove a lemma which gives the bounds on the cooperative throughput that can be achieved in this case.
\begin{lemma} \label{lemma1}
The cooperative throughput of a Type $\mathcal{C}$ link making use of a Tier 1 helper can be bounded as
\begin{subequations} \label{TC1Coop}
\begin{gather}
 	\mathcal{L}_\textnormal{Coop}^{\mathcal{C},1} \le \mathcal{T}_\textnormal{Coop}^{\mathcal{C},1}  \le  \mathcal{U}_\textnormal{Coop}^{\mathcal{C},1} \\
 	\shortintertext{where}
	\mathcal{L}_\textnormal{Coop}^{\mathcal{C},1} \triangleq \mathbb{G}(48.2,48.2)\times 5.5~\textnormal{(Mbps)} \\
 	\mathcal{U}_\textnormal{Coop}^{\mathcal{C},1} \triangleq \mathbb{G}\Big(\frac{r_k}{2},\frac{r_k}{2}\Big)\times 5.5~\textnormal{(Mbps)}.
\end{gather}
\end{subequations}
\end{lemma}
\begin{IEEEproof}
The proof of Lemma \ref{lemma1} is given in Appendix \ref{appendix1}.
\end{IEEEproof}

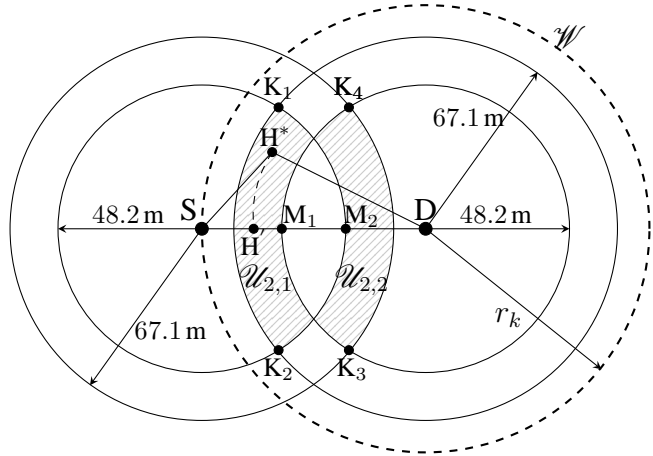
\begin{figure}[!t]
\centering
\begin{tikzpicture}[>=stealth, scale=0.85, every node/.style={transform shape}]
\tikzstyle{every node}=[draw,shape=circle, node distance=1cm];
\draw [fill =black](-1.75, 0) circle (0.1);
\draw [fill =black](1.75, 0) circle (0.1);
\node [draw=none](s) at (-1.95,0.3 ) {S};
\node [draw=none](d) at (1.75,0.3 ) {D};
\begin{scope}
\clip (1.75,0) circle (2.25);
\clip (-1.75,0) circle (3);
\clip (1.75,0) circle (2.25);
\fill[white] (-4,2.5)
rectangle (4,-2.5);
\pattern [pattern=north east lines,pattern color=gray!40]
 (-4,2.5)
rectangle (4,-2.5);
\end{scope}
\begin{scope}
\clip (-1.75,0) circle (2.25);
\clip (-1.75,0) circle (3);
\clip (1.75,0) circle (3);
\fill[white] (-4,2.5)
rectangle (4,-2.5);
\pattern [pattern=north east lines,pattern color=gray!40]
 (-4,2.5)
rectangle (4,-2.5);
\end{scope}
\begin{scope}
\clip (1.75,0) circle (2.25);
\clip (-1.75,0) circle (2.25);
\fill[white] (-4,2.5)
rectangle (4,-2.5);
\pattern [pattern=north east lines,pattern color=white]
 (-4,2.5)
rectangle (4,-2.5);
\end{scope}
\draw [dashed,thick](1.75, 0) circle (3.5);
\draw [fill =black](-0.5, 0) circle (0.07);
\draw [fill =black](0.5, 0) circle (0.07);
\draw [fill =black](-0.55, 1.9) circle (0.07);
\draw [fill =black](0.55, 1.9) circle (0.07);
\draw [fill =black](-0.55, -1.9) circle (0.07);
\draw [fill =black](0.55, -1.9) circle (0.07);
\draw [fill =black](-0.65, 1.2) circle (0.07);
\draw [fill =black](-0.94, 0) circle (0.07);
\draw (-1.75, 0) circle (2.25);

\draw  (1.75,0) circle (3);
\draw  (-1.75,0) circle (3);
\draw (1.75, 0) circle (2.25);
\node [draw=none] (v1) at (0.75,0.2) {\small${\text{M}_2}$};
\node [draw=none] (v1) at (-0.2,0.2) {\small${\text{M}_1}$};
\node [draw=none] (v1) at (-0.55,-2.2) {\small${\text{K}_2}$};
\node [draw=none] (v1) at (0.6,-2.2) {\small${\text{K}_3}$};
\node [draw=none] (v1) at (-0.55,2.2) {\small${\text{K}_1}$};
\node [draw=none] (v1) at (0.55,2.2) {\small${\text{K}_4}$};
\node [draw=none] (v1) at (0.55,2.2) {\small${\text{K}_4}$};
\node [draw=none] (v1) at (-0.6,1.45) {\small${\text{H}^*}$};
\node [draw=none] (v1) at (-0.93,-0.2) {\small${\text{H}^{'}}$};
\node [draw=none] (v1) at (-0.75,-0.8) {$\mathscr{U}_{2,1}$};
\node [draw=none] (v1) at (0.75,-0.8) {$\mathscr{U}_{2,2}$};
\node [draw=none] (v1) at (4,3) {$\mathscr{W}$};
\draw [->] (1.75,0) -- node [draw=none,  text width=1cm, pos=0.5, yshift=-0.5cm ,above, fill=none, node distance=2cm] {\small$48.2$\,m} (4,0);
\draw [->] (-1.75,0) -- node [draw=none,  text width=1cm, pos=0.5, yshift=-0.5cm ,above, fill=none, node distance=1cm] {\small $48.2$\,m} (-4,0);
\draw [->] (1.75,0) -- node [draw=none,  text width=1cm, pos=0.5,xshift=-0.15cm, yshift=-0.25cm ,above, fill=none, node distance=2cm] {\small$67.1$\,m} (3.5,2.45);
\draw [->] (-1.75,0) -- node [draw=none,  text width=1cm, pos=0.5,xshift=0.35cm, yshift=-1cm ,above, fill=none, node distance=1cm] {\small $67.1$\,m} (-3.5,-2.45);
\draw [->] (1.75,0) -- node [draw=none,  text width=0.5cm, pos=0.5, yshift=-0.25cm ,midway, fill=none, node distance=1cm] {$r_k$} (4.5,-2.2);
\draw (-1.75,0)--(1.75,0);
\draw (-0.65,1.2)--(1.75,0);
\draw (-0.65,1.2)--(-1.75,0);
\draw [dashed] (-0.65,1.2) .. controls (-1,0.65) and (-0.95,0.2) .. (-0.94,0);
\end{tikzpicture}

\caption{A typical Type $\mathcal{C}$ link making use of a Tier 2 helper, $67.1\le r_k\le74.7$.}
\label{fig3}
\end{figure}
\subsection{The Cooperative Throughput Using a Tier 2 Helper}  \label{SecTier2}
Fig. \ref{fig3} shows a typical Type $\mathcal{C}$ link making use of a Tier 2 helper as shown in Table \ref{T2}.
The probability that cooperation is made through a Tier 2 helper, referred to as $\mathcal{P}_{\mathcal{C},2}$, equals the probability that there is no helper in $\mathscr{U}_1$, and at least one helper is located in $\mathscr{U}_2 = \mathscr{U}_{2,1}\cup\mathscr{U}_{2,2}$. Using the fact that there are exactly $k-1$ nodes in a circle centered at D with radius $r_k$ ($\mathscr{W}$ in Fig. \ref{fig3}) one can obtain
\begin{align} \notag
\mathcal{P}_{\mathcal{C},2}&= \Pr\{\mathbb{N}(\mathscr{U}_2)\ge1 \text{ and } \mathbb{N}(\mathscr{U}_1)=0\} \\
&= \Pr\{\mathbb{N}(\mathscr{U}_1)=0\}-\Pr\{\mathbb{N}(\mathscr{U}_1)=0 \text{ and }\mathbb{N}(\mathscr{U}_2)=0 \} \notag \\
& = \Pr\{\mathbb{N}(\mathscr{W}\!-\!\mathscr{U}_1)=k-1\}-\Pr\{\mathbb{N}(\mathscr{W}\!-\!(\mathscr{U}_1\cup\mathscr{U}_2))=k-1\} \notag \\
& = \bigg(\frac{\mathcal{S}(\mathscr{W})\!-\!\mathcal{S}(\mathscr{U}_{1})}{\mathcal{S}(\mathscr{W})}\bigg)^{k-1}-\bigg(\frac{\mathcal{S}(\mathscr{W})\!-\!(\mathcal{S}(\mathscr{U}_1)+\mathcal{S}(\mathscr{U}_2))}{\mathcal{S}(\mathscr{W})}\bigg)^{k-1} \notag\\  \label{s1}
& = \bigg(1-\frac{\mathcal{S}(\mathscr{U}_1)}{\pi\,\!r_k^2}\bigg)^{\!k-1}-\bigg(1-\frac{\mathcal{S}(\mathscr{U}_1)+\mathcal{S}(\mathscr{U}_2)}{\pi\,\!r_k^2}\bigg)^{\!k-1}
\end{align}
where $\mathcal{S}(\mathscr{U}_2) = 2\big(\mathbb{A}(67.1,48.2,r_k)-\mathbb{A}(48.2,48.2,r_k)\big)$, and $\mathbb{A}(r_1,r_2,\ell)$ was defined in \eqref{overlap} and \eqref{overlaph}.
\par We now use a procedure similar to that presented in Lemma \ref{lemma1} to obtain the maximum and minimum of $\PcoopsCb$.  This procedure is summarized in the following lemma.
\begin{lemma} \label{lemma2}
For a Type $\mathcal{C}$ link making use of a Tier 2 helper the cooperative throughput is bounded as
\begin{subequations}\label{TC2Coop}
\begin{gather}
 	\mathcal{L}_\textnormal{Coop}^{\mathcal{C},2} \le \mathcal{T}_\textnormal{Coop}^{\mathcal{C},2}  \le  \mathcal{U}_\textnormal{Coop}^{\mathcal{C},2} \\
 	\shortintertext{where}
	\mathcal{L}_\textnormal{Coop}^{\mathcal{C},2} \triangleq \mathbb{G}(48.2,67.1)\times 3.67~\textnormal{(Mbps)} \\
 	\mathcal{U}_\textnormal{Coop}^{\mathcal{C},2} \triangleq \mathbb{G}(48.2,r_k-48.2)\times 3.67~\textnormal{(Mbps)}.
\end{gather}
\end{subequations}
\end{lemma}
\begin{IEEEproof}
The proof of Lemma \ref{lemma2} is given in Appendix \ref{appendix2}.
\end{IEEEproof}
\subsection{The Cooperative Throughput Using a Tier 3 Helper} \label{SecTier3}
A typical Type $\mathcal{C}$ link making use of a Tier 3 helper is illustrated in Fig. \ref{fig4}.  A Tier 3 helper can be located either in $\mathscr{U}_{3,1}$ or in  $\mathscr{U}_{3,2}$ (the shaded areas in Fig \ref{fig4}).  These areas are characterized as
\begin{subequations}
\begin{align}
48.2\le \dsh &\le 67.1\\
48.2\le \dhd &\le 67.1\\
r_k &\le \dsh+\dhd.
\end{align}
\end{subequations}
A Tier 3 helper is used for cooperation when there is no helper in $\mathscr{U}_1$ and $\mathscr{U}_2$, and there exists at least one helper in $\mathscr{U}_3\triangleq\mathscr{U}_{3,1}\cup\mathscr{U}_{3,2}$.
The probability of this event, referred to as $\mathcal{P}_{\mathcal{C},3}$, can be obtained
\begin{align} \notag
\mathcal{P}_{\mathcal{C},3}&= \Pr\{\mathbb{N}(\mathscr{U}_3)\ge1 \text{ and }\mathbb{N}(\mathscr{U}_1)=0 \text{ and }\mathbb{N}(\mathscr{U}_2)=0\} \\
&= \Pr\{\mathbb{N}(\mathscr{U}_1)=0 \text{ and }\mathbb{N}(\mathscr{U}_2)=0 \}
-\Pr\{\mathbb{N}(\mathscr{U}_1)=0 \text{ and }\mathbb{N}(\mathscr{U}_2)=0 \text{ and }\mathbb{N}(\mathscr{U}_3)=0\}\notag \\
& = \Pr\{\mathbb{N}\big(\mathscr{W}\!-\!(\mathscr{U}_1\cup\mathscr{U}_2)\big)=k-1\}
-\Pr\{\mathbb{N}\big(\mathscr{W}\!-\!(\mathscr{U}_1\cup\mathscr{U}_2\cup\mathscr{U}_3)\big)=k-1\} \notag \\
& = \bigg(\frac{\mathcal{S}(\mathscr{W})\!-\!(\mathcal{S}(\mathscr{U}_{1})+\mathcal{S}(\mathscr{U}_{2}))}{\mathcal{S}(\mathscr{W})}\bigg)^{k-1}-\bigg(\frac{\mathcal{S}(\mathscr{W})\!-\!(\mathcal{S}(\mathscr{U}_1)+\mathcal{S}(\mathscr{U}_2)+\mathcal{S}(\mathscr{U}_{3}))}{\mathcal{S}(\mathscr{W})}\bigg)^{k-1} \notag\\  \label{s1}
& = \bigg(1-\frac{\mathcal{S}(\mathscr{U}_1)+\mathcal{S}(\mathscr{U}_2)}{\pi\,\!r_k^2}\bigg)^{\!k-1}-\bigg(1-\frac{\mathcal{S}(\mathscr{U}_1)+\mathcal{S}(\mathscr{U}_2)+\mathcal{S}(\mathscr{U}_3)}{\pi\,\!r_k^2}\bigg)^{\!k-1}
\end{align}
where $\mathcal{S}(\mathscr{U}_3) = \mathbb{A}(67.1,67.1,r_k)-2\mathbb{A}(67.1,48.2,r_k)+\mathbb{A}(48.2,48.2,r_k)$, and $\mathbb{A}(r_1,r_2,\ell)$ was defined in \eqref{overlap} and \eqref{overlaph}.
The maximum and minimum of $\PcoopsCc$ in this case are easy to obtain.  Indeed, using eq. \eqref{PSuccessDef} along with the fact that $\mathbb{Q}(x)$ is a monotonically decreasing function of $x$, we can readily see that
\begin{equation}
\mathbb{G}(67.1,67.1) \le \mathcal{P}^{\text{Succ},\mathcal{C},3}_{\text{Coop}} \le  \mathbb{G}(48.2,48.2).
\end{equation}
Hence, the cooperative throughput of a Type $\mathcal{C}$ link that utilizes a Tier 3 helper, i.e., $\mathcal{T}_\text{Coop}^{\mathcal{C},3}$, can be bounded as
\begin{subequations}
\begin{gather}
 	\mathcal{L}_\text{Coop}^{\mathcal{C},3} \le \mathcal{T}_\text{Coop}^{\mathcal{C},3}  \le  \mathcal{U}_\text{Coop}^{\mathcal{C},3} \\
	\mathcal{L}_\text{Coop}^{\mathcal{C},3} \triangleq \mathbb{G}(67.1,67.1)\times 2.75~\text{(Mbps)} \\
 	\mathcal{U}_\text{Coop}^{\mathcal{C},3} \triangleq \mathbb{G}(48.2,48.2)\times 2.75~\text{(Mbps)}.
\end{gather}
\end{subequations}
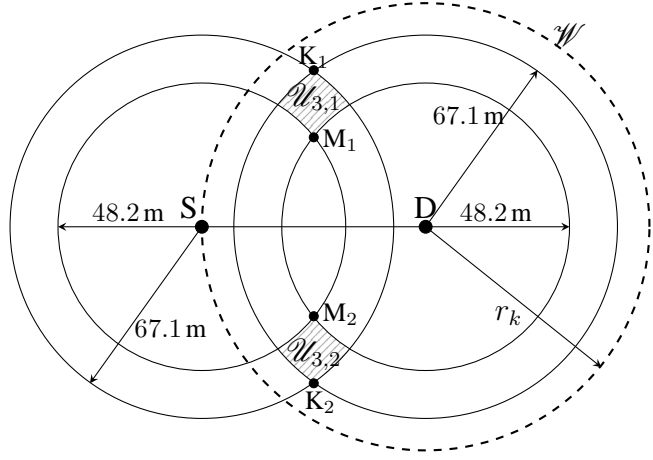
\begin{figure}[!t]
\centering
\begin{tikzpicture}[>=stealth, scale=0.85, every node/.style={transform shape}]
\tikzstyle{every node}=[draw,shape=circle, node distance=1cm];
\draw [fill =black](-1.75, 0) circle (0.1);
\draw [fill =black](1.75, 0) circle (0.1);
\node [draw=none](s) at (-1.95,0.3 ) {S};
\node [draw=none](d) at (1.75,0.3 ) {D};

\begin{scope}
\clip (1.75,0) circle (2.25);
\clip (-1.75,0) circle (3);
\clip (1.75,0) circle (2.25);
\fill[white] (-4,2.5)
rectangle (4,-2.5);
\pattern [pattern=north east lines,pattern color=white]
 (-4,2.5)
rectangle (4,-2.5);
\end{scope}
\begin{scope}
\clip (-1.75,0) circle (3);
\clip (1.75,0) circle (3);
\fill[white] (-4,2.5)
rectangle (4,-2.5);
\pattern [pattern=north east lines,pattern color=gray!60]
 (-4,2.5)
rectangle (4,-2.5);
\end{scope}
\begin{scope}
\clip (-1.75,0) circle (2.25);
\clip (-1.75,0) circle (3);
\clip (1.75,0) circle (3);
\fill[white] (-4,2.5)
rectangle (4,-2.5);
\pattern [pattern=north east lines,pattern color=white]
 (-4,2.5)
rectangle (4,-2.5);
\end{scope}
\begin{scope}

\clip (1.75,0) circle (2.25);
\clip (-1.75,0) circle (3);
\clip (1.75,0) circle (3);
\fill[white] (-4,2.5)
rectangle (4,-2.5);
\pattern [pattern=north east lines,pattern color=white]
 (-4,2.5)
rectangle (4,-2.5);
\end{scope}
\begin{scope}
\clip (1.75,0) circle (2.25);
\clip (-1.75,0) circle (2.25);
\fill[white] (-4,2.5)
rectangle (4,-2.5);
\pattern [pattern=north east lines,pattern color=white]
 (-4,2.5)
rectangle (4,-2.5);
\end{scope}
\draw (-1.75, 0) circle (2.25);
\draw [dashed,thick](1.75, 0) circle (3.5);
\draw  (1.75,0) circle (3);
\draw  (-1.75,0) circle (3);
\draw (1.75, 0) circle (2.25);
\draw [fill =black](0, 1.4) circle (0.07);
\draw [fill =black](0, -1.4) circle (0.07);
\draw [fill =black](0, 2.45) circle (0.07);
\draw [fill =black](0, -2.45) circle (0.07);
\node [draw=none] (v1) at (0,-2) {$\mathscr{U}_{3,2}$};
\node [draw=none] (v1) at (0,2) {$\mathscr{U}_{3,1}$};
\node [draw=none] (v1) at (0,2.65) {\small${\text{K}_1}$};
\node [draw=none] (v1) at (0.1,-2.75) {\small${\text{K}_2}$};
\node [draw=none] (v1) at (0.42,1.35) {\small${\text{M}_1}$};
\node [draw=none] (v1) at (0.42,-1.37) {\small${\text{M}_2}$};
\draw [->] (1.75,0) -- node [draw=none,  text width=1cm, pos=0.5, yshift=-0.5cm ,above, fill=none, node distance=2cm] {\small$48.2$\,m} (4,0);
\draw [->] (-1.75,0) -- node [draw=none,  text width=1cm, pos=0.5, yshift=-0.5cm ,above, fill=none, node distance=1cm] {\small $48.2$\,m} (-4,0);
\draw [->] (1.75,0) -- node [draw=none,  text width=1cm, pos=0.5, xshift=-0.15cm,yshift=-0.25cm ,above, fill=none, node distance=2cm] {\small$67.1$\,m} (3.5,2.45);
\draw [->] (-1.75,0) -- node [draw=none,  text width=1cm, pos=0.5,xshift=0.35cm, yshift=-1cm ,above, fill=none, node distance=1cm] {\small $67.1$\,m} (-3.5,-2.45);
\draw [->] (1.75,0) -- node [draw=none,  text width=0.5cm, pos=0.5, yshift=-0.25cm ,midway, fill=none, node distance=1cm] {$r_k$} (4.5,-2.2);
\draw (-1.75,0)--(1.75,0);
\node [draw=none] (v1) at (4,3) {$\mathscr{W}$};
\end{tikzpicture}
\caption{A typical Type $\mathcal{C}$ link making use of a Tier 3 helper, $67.1\le r_k\le74.7$.}
\label{fig4}
\end{figure}
Using the results given in Subsections \ref{SecTier1} through \ref{SecTier3}, one can readily see
\begin{subequations}
\begin{gather}
    \mathcal{L^C}(r_k)\le\mathcal{T^C}(r_k)\le\mathcal{U^C}(r_k)\\
    \shortintertext{where}
	\mathcal{L}^{\mathcal{C}}(r_k)=\sum_{i=1}^3\mathcal{P}_{\mathcal{C},i}\,\mathcal{L}_\text{Coop}^{\mathcal{C},i}+\bigg(1-\sum_{i=1}^3\mathcal{P}_{\mathcal{C},i}\bigg)\Ps(r_k)\times 2~\text{(Mbps)} \\
	\mathcal{U}^{\mathcal{C}}(r_k)=\sum_{i=1}^3\mathcal{P}_{\mathcal{C},i}\,\mathcal{U}_\text{Coop}^{\mathcal{C},i}+\bigg(1-\sum_{i=1}^3\mathcal{P}_{\mathcal{C},i}\bigg)\Ps(r_k)\times 2~\text{(Mbps)}.
\end{gather}
\end{subequations}
In consequence, the average throughput of a Type $\mathcal{C}$ link can be bounded as
\begin{subequations}\label{TC}
\begin{gather}
    \mathcal{\overline L^C}\le\mathcal{\overline T^C}\le\mathcal{\overline U^C}\\
    \shortintertext{where}
	\mathcal{\overline L^C}=\int_{67.1}^{74.7} 	\mathcal{L^C}(r) \frk \mathrm{d}r  \\
	\mathcal{\overline U^C}=\int_{67.1}^{74.7} 	\mathcal{U^C}(r) \frk \mathrm{d}r.
\end{gather}
\end{subequations}
\begin{figure*}[!t]
\centering
\begin{tikzpicture}[>=stealth]
\tikzstyle{every node}=[draw,shape=circle, node distance=1cm];
\draw [fill =black](-2.25, 0) circle (0.1);
\draw [fill =black](2.25, 0) circle (0.1);
\node [draw=none](s) at (-2.45,0.3 ) {S};
\node [draw=none](d) at (2.25,0.3 ) {D};

\begin{scope}
\clip (2.25,0) circle (2.5);
\clip (-2.25,0) circle (3.5);
\clip (2.25,0) circle (2.5);
\fill[white] (-4,2.5)
rectangle (4,-2.5);
\pattern [pattern=north east lines,pattern color=white]
 (-4,2.5)
rectangle (4,-2.5);
\end{scope}
\begin{scope}
\clip (2.25,0) circle (3.5);
\clip (-2.25,0) circle (4);
\clip (2.25,0) circle (4);
\fill[white] (-4.5,3)
rectangle (4.5,-3);
\pattern [pattern=crosshatch dots,pattern color=gray!50]
 (-4.5,3)
rectangle (4.5,-3);
\end{scope}
\begin{scope}
\clip (-2.25,0) circle (3.5);
\clip (-2.25,0) circle (4);
\clip (2.25,0) circle (4);
\fill[white] (-4.5,3)
rectangle (4.5,-3);
\pattern [pattern=crosshatch dots,pattern color=gray!50]
 (-4.5,3)
rectangle (4.5,-3);
\end{scope}
\begin{scope}
\clip (-2.25,0) circle (3.5);
\clip (2.25,0) circle (3.5);
\fill[white] (-4.5,3)
rectangle (4.5,-3);
\pattern [pattern=horizontal lines,pattern color=gray!50]
 (-4.5,3)
rectangle (4.5,-3);
\end{scope}
\begin{scope}
\clip (2.25,0) circle (2.5);
\clip (-2.25,0) circle (4);
\clip (2.25,0) circle (4);
\fill[white] (-4.5,3)
rectangle (4.5,-3);
\pattern [pattern=vertical lines,pattern color=gray!80]
 (-4.5,3)
rectangle (4.5,-3);
\end{scope}
\begin{scope}
\clip (-2.25,0) circle (2.5);
\clip (-2.25,0) circle (4);
\clip (2.25,0) circle (4);
\fill[white] (-4.5,3)
rectangle (4.5,-3);
\pattern [pattern=vertical lines,pattern color=gray!80]
 (-4.5,3)
rectangle (4.5,-3);
\end{scope}
\begin{scope}
\clip (2.25,0) circle (2.5);
\clip (-2.25,0) circle (4);
\clip (2.25,0) circle (4);
\fill[white] (-4.5,3)
rectangle (4.5,-3);
\pattern [pattern=vertical lines,pattern color=gray!80]
 (-4.5,3)
rectangle (4.5,-3);
\end{scope}
\begin{scope}
\clip (2.25,0) circle (2.5);
\clip (-2.25,0) circle (3.5);
\clip (2.25,0) circle (3.5);
\fill[white] (-4.5,3)
rectangle (4.5,-3);
\pattern [pattern=grid,pattern color=black!30]
 (-4.5,3)
rectangle (4.5,-3);
\end{scope}
\begin{scope}
\clip (-2.25,0) circle (2.5);
\clip (-2.25,0) circle (3.5);
\clip (2.25,0) circle (3.5);
\fill[white] (-4.5,3)
rectangle (4.5,-3);
\pattern [pattern=grid,pattern color=black!30]
 (-4.5,3)
rectangle (4.5,-3);
\end{scope}

\begin{scope}
\clip (2.25,0) circle (2.5);
\clip (-2.25,0) circle (3.5);
\clip (2.25,0) circle (3.5);
\fill[white] (-4.5,3)
rectangle (4.5,-3);
\pattern [pattern=grid,pattern color=black!30]
 (-4.5,3)
rectangle (4.5,-3);
\end{scope}
\begin{scope}
\clip (2.25,0) circle (2.5);
\clip (-2.25,0) circle (2.5);
\fill[white] (-4.5,3)
rectangle (4.5,-3);
\pattern [pattern=north east lines,pattern color=gray]
 (-4.5,3)
rectangle (4.5,-3);
\end{scope}
\draw [pattern=north east lines](7.5, 1.5) rectangle (8,2);
\draw [pattern=grid,pattern color=black!30](7.5, 0.75) rectangle (8,1.25);
\draw [pattern=horizontal lines,pattern color=gray!80](7.5, 0.5) rectangle (8,0);
\draw  [pattern=vertical lines,pattern color=gray!80](7.5,- 0.75) rectangle (8,-0.25);
\draw  [pattern=crosshatch dots,pattern color=gray](7.5,- 1.5) rectangle (8,-1);
\node [draw=none] (v1) at (7.2,1.63) {\small$\mathscr{V}_1$};
\node [draw=none] (v1) at (7.2,0.9) {\small$\mathscr{V}_2$};
\node [draw=none] (v1) at (7.2,0.15) {\small$\mathscr{V}_3$};
\node [draw=none] (v1) at (7.2,-0.6) {\small$\mathscr{V}_4$};
\node [draw=none] (v1) at (7.2,-1.35) {\small$\mathscr{V}_5$};
\draw (-2.25, 0) circle (2.5);
\draw (-2.25, 0) circle (4);
\draw [dashed,thick](2.25, 0) circle (4.5);
\draw  (2.25,0) circle (3.5);
\draw  (-2.25,0) circle (3.5);
\draw (2.25, 0) circle (2.5);
\draw (2.25, 0) circle (4);
\draw [fill =black](0, 0) circle (0.07);
\draw [fill =black](0.25, 0) circle (0.07);
\draw [fill =black](-0.25,0) circle (0.07);
\draw [fill =black](1.25,0) circle (0.07);
\draw [fill =black](-1.25,0) circle (0.07);
\draw [fill =black](-0.68,1.94) circle (0.07);
\draw [fill =black](-0.68,-1.94) circle (0.07);
\draw [fill =black](0.68,1.94) circle (0.07);
\draw [fill =black](0.68,-1.94) circle (0.07);
\draw [fill =black](0.44,2.98) circle (0.07);
\draw [fill =black](-0.44,2.98) circle (0.07);
\draw [fill =black](0.44,-2.98) circle (0.07);
\draw [fill =black](-0.44,-2.98) circle (0.07);
\draw [fill =black](-0.68,1.94) circle (0.07);
\draw [fill =black](-0.68,-1.94) circle (0.07);
\draw [fill =black](0.68,1.94) circle (0.07);
\draw [fill =black](0.68,-1.94) circle (0.07);
\draw [fill =black](1.11,2.2) circle (0.07);
\draw [fill =black](-1.11,2.2) circle (0.07);
\draw [fill =black](1.11,-2.2) circle (0.07);
\draw [fill =black](-1.11,-2.2) circle (0.07);
\draw [fill =black](0, 1.1) circle (0.07);
\draw [fill =black](0, -1.1) circle (0.07);
\draw [fill =black](0, 2.7) circle (0.07);
\draw [fill =black](0, -2.7) circle (0.07);
\node [draw=none] (v1) at (0,0.2) {\small$\text{o}$};
\node [draw=none] (v1) at (0.45,-0.2) {\small${\text{b}_1}$};
\node [draw=none] (v1) at (-0.45,-0.2) {\small${\text{b}_2}$};
\node [draw=none] (v1) at (1.45,-0.2) {\small${\text{c}_1}$};
\node [draw=none] (v1) at (-1.45,-0.2) {\small${\text{c}_2}$};
\node [draw=none] (v1) at (0,1.4) {\small${\text{d}_1}$};
\node [draw=none] (v1) at (0,-1.4) {\small${\text{d}_2}$};
\node [draw=none] (v1) at (0.41,1.94) {\small${\text{e}_1}$};
\node [draw=none] (v1) at (-0.41,1.94) {\small${\text{e}_2}$};
\node [draw=none] (v1) at (0.41,-1.94) {\small${\text{e}_3}$};
\node [draw=none] (v1) at (-0.41,-1.94) {\small${\text{e}_4}$};
\node [draw=none] (v1) at (1.11,2.5) {\small${\text{f}_1}$};
\node [draw=none] (v1) at (-1.11,2.5) {\small${\text{f}_2}$};
\node [draw=none] (v1) at (1.11,-2.5) {\small${\text{f}_3}$};
\node [draw=none] (v1) at (-1.11,-2.5) {\small${\text{f}_4}$};
\node [draw=none] (v1) at (0,3) {\small${\text{g}_1}$};
\node [draw=none] (v1) at (0,-3) {\small${\text{g}_2}$};
\node [draw=none] (v1) at (0.44,3.3) {\small${\text{i}_1}$};
\node [draw=none] (v1) at (-0.44,3.3) {\small${\text{i}_2}$};
\node [draw=none] (v1) at (0.44,-3.3) {\small${\text{i}_3}$};
\node [draw=none] (v1) at (-0.44,-3.3) {\small${\text{i}_4}$};
\draw [->] (2.25,0) -- node [draw=none,  text width=1cm, pos=0.5, yshift=-0.5cm ,above, fill=none, node distance=2cm] {\small$48.2$\,m} (4.75,0);
\draw [->] (-2.25,0) -- node [draw=none,  text width=1cm, pos=0.5, yshift=-0.5cm ,above, fill=none, node distance=1cm] {\small $48.2$\,m} (-4.75,0);
\draw [->] (2.25,0) -- node [draw=none,  text width=1cm, pos=0.5,xshift=-0.23cm, yshift=-0.25cm ,above, fill=none, node distance=2cm] {\small$67.1$\,m} (4,3);
\draw [->] (-2.25,0) -- node [draw=none,  text width=1cm, pos=0.5,xshift=0.35cm, yshift=-1cm ,above, fill=none, node distance=1cm] {\small $67.1$\,m} (-4,-3);
\draw [->] (2.25,0) -- node [draw=none,  text width=1cm, pos=0.5, xshift=-0.44cm,yshift=-2.35cm ,above, fill=none, node distance=2cm] {\small$74.7$\,m} (2.25,-4);
\draw [->] (-2.25,0) -- node [draw=none,  text width=1cm, pos=0.5, xshift=-0.44cm ,above, fill=none, node distance=1cm] {\small $74.7$\,m} (-2.25,4);
\draw [->] (2.25,0) -- node [draw=none,  text width=0.5cm, pos=0.5, yshift=-0.15cm ,midway, fill=none, node distance=1cm] {$r_k$} (5.6,-3);
\draw (-2.25,0)--(2.25,0);
\end{tikzpicture}
\caption{A typical Type $\mathcal{D}$ link with $74.7< r_k\le96.4$ and the operating regions of Tiers 1 through 5 helpers.}
\label{fig5}
\end{figure*}
\section{Bounds on The Average Throughput of a Type $\mathcal{D}$ Link} \label{boundsD}
In this section, we use the analysis given in Section \ref{boundsC} to obtain the maximum and minimum cooperative throughputs of a Type $\mathcal{D}$ link.
As mentioned in Subsection \ref{sysmodD}, a Type $\mathcal{D}$ link whose source and destination are more than $96.4$ meters apart (i.e., $96.4<r_k\le 100$), can not use a Tier 1 helper.  Therefore, we divide our analysis into two parts, viz., $74.7< r_k\le96.4$ and $96.4< r_k\le100$.
Recall that in our helper selection algorithm, the lower the tier index, the higher the selection priority.  Hence, using the procedure outlined in Section \ref{boundsC} for evaluating $\mathcal{P}_{\mathcal{C},1}$ through $\mathcal{P}_{\mathcal{C},3}$, we can obtain the probability that a Type $\mathcal{D}$ link chooses a Tier $i$ helper as
\begin{align}
\mathcal{P}_{\mathcal{D},i} = \begin{cases}
               1-\Big(1-\frac{\mathcal{S}(\mathscr{V}_1)}{\mathcal{S}(\mathscr{W})}\Big)^{k-1}, & i=1 \\
               \Big(1-\frac{\sum_{\ell=1}^{i-1}\mathcal{S}(\mathscr{V}_\ell)}{\mathcal{S}(\mathscr{W})}\Big)^{k-1}-\Big(1-\frac{\sum_{\ell=1}^{i}\mathcal{S}(\mathscr{V}_\ell)}{\mathcal{S}(\mathscr{W})}\Big)^{k-1}, & i=2,\ldots,5
               \end{cases}
\end{align}
where $\mathcal{S}(\mathscr{W})=\pi r_k^2$ and $\mathcal{S}(\mathscr{V}_i)$ depends on $r_k$ and should be evaluated for $74.7< r_k\le96.4$ and $96.4\le r_k\le100$, separately.  For the case where $74.7< r_k\le96.4$, we can readily see from Fig. \ref{fig5} and \eqref{overlap} that
\begin{subequations}
\begin{align}
&\mathcal{S}(\mathscr{V}_1)= \mathbb{A}(48.2,48.2,r_k), \\ \label{SV2}
&\mathcal{S}(\mathscr{V}_2) = 2\big(\mathbb{A}(48.2,67.1,r_k) - \mathcal{S}(\mathscr{V}_1)\big),   \\\label{SV3}
&\mathcal{S}(\mathscr{V}_3) = \mathbb{A}(67.1,67.1,r_k) -\mathcal{S}(\mathscr{V}_2) - \mathcal{S}(\mathscr{V}_1),  \\\label{SV4}
&\mathcal{S}(\mathscr{V}_4) = 2\big(\mathbb{A}(48.2,74.7,r_k) - \mathcal{S}(\mathscr{V}_1)\big) - \mathcal{S}(\mathscr{V}_2),   \\\label{SV5}
&\mathcal{S}(\mathscr{V}_5) = 2\big(\mathbb{A}(67.1,74.7,r_k) -\mathbb{A}(67.1,67.1,r_k)\big) - \mathcal{S}(\mathscr{V}_4).
\end{align}
\end{subequations}
\begin{figure*}[!t]
\centering
\begin{tikzpicture}[>=stealth, scale=1]
\tikzstyle{every node}=[draw,shape=circle, node distance=1cm];
\draw [fill =black](-2.65, 0) circle (0.1);
\draw [fill =black](2.65, 0) circle (0.1);
\node [draw=none](s) at (-2.85,0.3 ) {S};
\node [draw=none](d) at (2.65,0.3 ) {D};
\begin{scope}
\clip (2.65,0) circle (2.5);
\clip (-2.65,0) circle (3.5);
\clip (2.65,0) circle (2.5);
\fill[white] (-4,2.5)
rectangle (4,-2.5);
\pattern [pattern=north east lines,pattern color=white]
 (-4,2.5)
rectangle (4,-2.5);
\end{scope}
\begin{scope}
\clip (2.65,0) circle (3.5);
\clip (-2.65,0) circle (4);
\clip (2.65,0) circle (4);
\fill[white] (-4.5,3)
rectangle (4.5,-3);
\pattern [pattern=crosshatch dots,pattern color=gray!50]
 (-4.5,3)
rectangle (4.5,-3);
\end{scope}
\begin{scope}
\clip (-2.65,0) circle (3.5);
\clip (-2.65,0) circle (4);
\clip (2.65,0) circle (4);
\fill[white] (-4.5,3)
rectangle (4.5,-3);
\pattern [pattern=crosshatch dots,pattern color=gray!50]
 (-4.5,3)
rectangle (4.5,-3);
\end{scope}
\begin{scope}
\clip (-2.65,0) circle (3.5);
\clip (2.65,0) circle (3.5);
\fill[white] (-4.5,3)
rectangle (4.5,-3);
\pattern [pattern=north east lines,pattern color=gray]
 (-4.5,3)
rectangle (4.5,-3);
\end{scope}
\begin{scope}
\clip (2.65,0) circle (2.5);
\clip (-2.65,0) circle (4);
\clip (2.65,0) circle (4);
\fill[white] (-4.5,3)
rectangle (4.5,-3);
\pattern [pattern=vertical lines,pattern color=gray!60]
 (-4.5,3)
rectangle (4.5,-3);
\end{scope}
\begin{scope}
\clip (-2.65,0) circle (2.5);
\clip (-2.65,0) circle (4);
\clip (2.65,0) circle (4);
\fill[white] (-4.5,3)
rectangle (4.5,-3);
\pattern [pattern=vertical lines,pattern color=gray!60]
 (-4.5,3)
rectangle (4.5,-3);
\end{scope}
\begin{scope}
\clip (2.65,0) circle (2.5);
\clip (-2.65,0) circle (4);
\clip (2.65,0) circle (4);
\fill[white] (-4.5,3)
rectangle (4.5,-3);
\pattern [pattern=vertical lines,pattern color=gray!60]
 (-4.5,3)
rectangle (4.5,-3);
\end{scope}
\begin{scope}
\clip (2.65,0) circle (2.5);
\clip (-2.65,0) circle (3.5);
\clip (2.65,0) circle (3.5);
\fill[white] (-4.5,3)
rectangle (4.5,-3);
\pattern [pattern=grid,pattern color=gray!30]
 (-4.5,3)
rectangle (4.5,-3);
\end{scope}
\begin{scope}
\clip (-2.65,0) circle (2.5);
\clip (-2.65,0) circle (3.5);
\clip (2.65,0) circle (3.5);
\fill[white] (-4.5,3)
rectangle (4.5,-3);
\pattern [pattern=grid,pattern color=gray!30]
 (-4.5,3)
rectangle (4.5,-3);
\end{scope}

\begin{scope}
\clip (2.65,0) circle (2.5);
\clip (-2.65,0) circle (3.5);
\clip (2.65,0) circle (3.5);
\fill[white] (-4.5,3)
rectangle (4.5,-3);
\pattern [pattern=grid,pattern color=gray!30]
 (-4.5,3)
rectangle (4.5,-3);
\end{scope}
\begin{scope}
\clip (2.65,0) circle (2.5);
\clip (-2.65,0) circle (2.5);
\fill[white] (-4.5,3)
rectangle (4.5,-3);
\pattern [pattern=north east lines,pattern color=gray!80]
 (-4.5,3)
rectangle (4.5,-3);
\end{scope}

\draw (-2.65, 0) circle (2.5);
\draw (-2.65, 0) circle (4);
\draw [dashed,thick](2.65, 0) circle (5.3);
\draw  (2.65,0) circle (3.5);
\draw  (-2.65,0) circle (3.5);
\draw (2.65, 0) circle (2.5);
\draw (2.65, 0) circle (4);
\draw [fill =black](0, 0) circle (0.07);
\draw [fill =black](0.17, 0) circle (0.07);
\draw [fill =black](-0.17,0) circle (0.07);
\draw [fill =black](0.85,0) circle (0.07);
\draw [fill =black](-0.85,0) circle (0.07);
\draw [fill =black](0.55,1.4) circle (0.07);
\draw [fill =black](-0.55,1.4) circle (0.07);
\draw [fill =black](0.55,-1.4) circle (0.07);
\draw [fill =black](-0.55,-1.4) circle (0.07);
\draw [fill =black](0, 2.3) circle (0.07);
\draw [fill =black](0, -2.3) circle (0.07);
\draw [fill =black](0.33, 2.64) circle (0.07);
\draw [fill =black](-0.33, 2.64) circle (0.07);
\draw [fill =black](0.33, -2.64) circle (0.07);
\draw [fill =black](-0.33, -2.64) circle (0.07);
\draw [fill =black](-0.9,1.8) circle (0.07);
\draw [fill =black](-0.9,-1.8) circle (0.07);
\draw [fill =black](0.9,-1.8) circle (0.07);
\draw [fill =black](0.9,1.8) circle (0.07);
\node [draw=none] (v1) at (0,0.2) {\small$\text{o}$};
\node [draw=none] (v1) at (0.37,-0.22) {\small${\text{b}_1}$};
\node [draw=none] (v1) at (-0.35,-0.2) {\small${\text{b}_2}$};
\node [draw=none] (v1) at (1.1,-0.2) {\small${\text{c}_1}$};
\node [draw=none] (v1) at (-1.05,-0.2) {\small${\text{c}_2}$};
\node [draw=none] (v1) at (0,2) {\small${\text{d}_1}$};
\node [draw=none] (v1) at (0,-2) {\small${\text{d}_2}$};
\node [draw=none] (v1) at (0.85,1.36) {\small${\text{e}_1}$};
\node [draw=none] (v1) at (-0.85,1.36) {\small${\text{e}_2}$};
\node [draw=none] (v1) at (-0.85,-1.36) {\small${\text{e}_3}$};
\node [draw=none] (v1) at (0.85,-1.36) {\small${\text{e}_4}$};
\node [draw=none] (v1) at (0.93,2.1) {\small${\text{f}_1}$};
\node [draw=none] (v1) at (-0.93,2.1) {\small${\text{f}_2}$};
\node [draw=none] (v1) at (-0.93,-2.1) {\small${\text{f}_3}$};
\node [draw=none] (v1) at (0.93,-2.1) {\small${\text{f}_4}$};
\node [draw=none] (v1) at (0.44,2.9) {\small${\text{i}_1}$};
\node [draw=none] (v1) at (-0.44,2.9) {\small${\text{i}_2}$};
\node [draw=none] (v1) at (0.44,-2.9) {\small${\text{i}_3}$};
\node [draw=none] (v1) at (-0.3,-2.9) {\small${\text{i}_4}$};
\draw [pattern=crosshatch dots,pattern color=gray!50](-0.75, -4.5) rectangle (-1.25,-5);
\draw [pattern=vertical lines,pattern color=gray!60](-2.5, -4.5) rectangle (-2,-5);
\draw  [pattern=north east lines,pattern color=gray!80](-3.75,- 4.5) rectangle (-3.25,-5);
\draw  [pattern=grid,pattern color=gray!30](-5,- 4.5) rectangle (-4.5,-5);
\node [draw=none] (v1) at (-5.25,-4.75) {\small$\mathscr{V}_2$};
\node [draw=none] (v1) at (-4,-4.75) {\small$\mathscr{V}_3$};
\node [draw=none] (v1) at (-2.75,-4.75) {\small$\mathscr{V}_4$};
\node [draw=none] (v1) at (-1.5,-4.75) {\small$\mathscr{V}_5$};
\draw [->] (2.65,0) -- node [draw=none,  text width=1cm, pos=0.5, yshift=-0.5cm ,above, fill=none, node distance=2cm] {\small$48.2$\,m} (5.15,0);
\draw [->] (-2.65,0) -- node [draw=none,  text width=1cm, pos=0.5, yshift=-0.5cm ,above, fill=none, node distance=1cm] {\small $48.2$\,m} (-5.15,0);
\draw [->] (2.65,0) -- node [draw=none,  text width=1cm, pos=0.5,xshift=-0.25cm, yshift=-0.25cm ,above, fill=none, node distance=2cm] {\small$67.1$\,m} (4.3,3.1);
\draw [->] (-2.65,0) -- node [draw=none,  text width=1cm, pos=0.5,xshift=0.35cm, yshift=-1cm ,above, fill=none, node distance=1cm] {\small $67.1$\,m} (-4.3,-3.1);
\draw [->] (2.65,0) -- node [draw=none,  text width=1cm, pos=0.5, xshift=-0.45cm,yshift=-2.35cm ,above, fill=none, node distance=2cm] {\small$74.7$\,m} (2.65,-4);
\draw [->] (-2.65,0) -- node [draw=none,  text width=1cm, pos=0.5, xshift=-0.45cm ,above, fill=none, node distance=1cm] {\small $74.7$\,m} (-2.65,4);
\draw [->] (2.65,0) -- node [draw=none,  text width=0.5cm, pos=0.5, yshift=-0.25cm ,midway, fill=none, node distance=1cm] {$r_k$} (6.67,-3.5);
\draw (-2.65,0)--(2.65,0);
\end{tikzpicture}
\caption{A typical Type $\mathcal{D}$ link with $96.4 < r_k\le100$ and the operating regions of Tiers 2 through 5 helpers.}
\label{fig6}
\end{figure*}
Recalling that there is no Tier 1 helper (or analogously $\mathscr{V}_1=\varnothing$) for the case where $96.4< r_k\le100$, and considering Fig. \ref{fig6}, we have
\begin{equation}
\mathcal{S}(\mathscr{V}_1) = 0.
\end{equation}
In this case, $\mathcal{S}(\mathscr{V}_i)$, $i=2, \ldots,5$ can be obtained from \eqref{SV2} through \eqref{SV5}, respectively.
In the next two subsections we obtain upper and lower bounds of the cooperative throughput for the cases where $74.7< r_k\le96.4$ and $96.4< r_k\le100$.
\subsection{Bounds on the Cooperative Throughput for $74.7< r_k\le 96.4$} \label{SubD1}
When $74.7< r_k\le 96.4$, a helper from each of the tiers shown in Table \ref{T3} (Tiers 1 through 5) can be used to increase the transmission rate between S and D nodes. 
Comparing Fig. \ref{fig5} with Figs. \ref{fig2}, \ref{fig3} and \ref{fig4}, one can readily see that the operating regions of Tier 1, 2 and 3 helpers for a Type $\mathcal{D}$ link are quite similar to those of Tier 1, 2 and 3 helpers for a Type $\mathcal{C}$ link, respectively.  Furthermore, the operating region of a Tier 4 helper for a Type $\mathcal{D}$ link is similar to that of a Tier 2 helper for a Type $\mathcal{C}$ link.  Note that the operating region of a Tier 5 helper for a Type $\mathcal{D}$ link is different from those of a Type $\mathcal{C}$ link.  However, using a procedure similar to that presented in Subsection \ref{SecTier3} for a Tier 3 helper, we can readily evaluate the maximum and minimum of the cooperative throughput in this case.  As a result, we can write
\begin{equation}
\lDai\le \mathcal{T}_\text{Coop}^{\mathcal{D}_1,i} \le \uDai, \quad i=1,\ldots,5
\end{equation}
where $\lDai$ and $\uDai$ ($i=1,\ldots,5$) are shown in Table \ref{T4} and $\mathcal{D}_1$ denotes a Type $\mathcal{D}$ link with $74.7<r_k\le96.4$.
Note also that Table \ref{T4} shows the locations of helpers in Fig. \ref{fig5} that can achieve the maximum and minimum $\Pcoops$ for each tier.
\begin{table*}[!t]
\caption{Illustration of the positions of the helpers achieving maximum and minimum $\Pcoops$ and their corresponding throughputs for a Type $\mathcal{D}$ link with $74.7< r_k\le96.4$.}
\centering
\renewcommand{\arraystretch}{1.25}
\begin{tabular}{|c|c|c|c|c|c|}
\hline
 \multicolumn{1}{|p{1.5cm}|}{\centering Helper's\\ Tier}&\multicolumn{1}{p{2.3cm}|}{\centering Point(s) with \\max. $\Pcoops$}&\multicolumn{1}{p{2.1cm}|}{\centering Points with\\ min. $\Pcoops$} &\raisebox{-1.5ex}[0pt]{$\lDai$ \text{(Mbps)}}&\raisebox{-1.5ex}[0pt]{$\uDai$ \text{(Mbps)}}\\ \hline
1 & \text o& $ \text d_1$, $ \text d_2$& $\mathbb{G}(48.2,48.2)\times 5.5$ & $\mathbb{G}\big(\frac{r_k}{2},\frac{r_k}{2}\big)\times 5.5$ \\ \hline
2\; &  $ \text b_1$, $ \text b_2$&\multicolumn{1}{p{2.1cm}|}{\centering e\textsubscript{1}, e\textsubscript{2}, e\textsubscript{3}, e\textsubscript{4}}&$\mathbb{G}(48.2,67.1)\times 3.67$ & $\mathbb{G}(48.2,r_k-48.2)\times 3.67$ \\ \hline
3\;& $ \text c_1$, $ \text c_2$&$ \text g_1$, $ \text g_2$& $\mathbb{G}(67.1,67.1)\times 2.75 $ & $\mathbb{G}(48.2,48.2)\times 2.75 $ \\ \hline
4\;& $ \text d_1$, $ \text d_2$&\multicolumn{1}{p{2cm}|}{\centering f\textsubscript{1}, f\textsubscript{2}, f\textsubscript{3}, f\textsubscript{4}}& $\mathbb{G}(48.2,74.7)\times 1.69 $& $\mathbb{G}(67.1,r_k-67.1)\times 1.69 $\\ \hline
5\;& \multicolumn{1}{p{2.2cm}|}{\centering e\textsubscript{1}, e\textsubscript{2}, e\textsubscript{3}, e\textsubscript{4}} &\multicolumn{1}{p{2cm}|}{\centering i\textsubscript{1}, i\textsubscript{2}, i\textsubscript{3}, i\textsubscript{4}}& $\mathbb{G}(67.1,74.7)\times 1.47$ &$\mathbb{G}(48.2,67.1)\times 1.47$\\ \hline
\end{tabular}
\label{T4}
\end{table*}
As a result, the cooperative throughput of a Type $\mathcal{D}$ link for $74.7< r_k \le96.4$ can be bounded as
\begin{subequations}
\begin{gather}
    \mathcal{L}^{\mathcal{D}_1}(r_k)\le\mathcal{T}^{\mathcal{D}_1}(r_k)\le\mathcal{U}^{\mathcal{D}_1}(r_k)\\
    \shortintertext{where}
	\mathcal{L}^{\mathcal{D}_1}(r_k)=\sum_{i=1}^5\mathcal{P}_{\mathcal{D},i}\,\mathcal{L}_\text{Coop}^{\mathcal{D}_1,i}+\bigg(1-\sum_{i=1}^5\mathcal{P}_{\mathcal{D},i}\bigg)\Ps(r_k)\times 1~\text{(Mbps)} \\
	\mathcal{U}^{\mathcal{D}_1}(r_k)=\sum_{i=1}^5\mathcal{P}_{\mathcal{D},i}\,\mathcal{U}_\text{Coop}^{\mathcal{D}_1,i}+\bigg(1-\sum_{i=1}^5\mathcal{P}_{\mathcal{D},i}\bigg)\Ps(r_k)\times 1~\text{(Mbps)}.
\end{gather}
\end{subequations}
\begin{table*}[!t]
\caption{Illustration of the positions of the helpers achieving maximum and minimum $\Pcoops$ and their corresponding throughputs for a Type $\mathcal{D}$ link with $96.4< r_k\le100$.}
\centering
\renewcommand{\arraystretch}{1.25}
\begin{tabular}{|c|c|c|c|c|c|}
\hline
 \multicolumn{1}{|p{1.5cm}|}{\centering Helper's\\ Tier}&\multicolumn{1}{p{2.3cm}|}{\centering Point(s) with \\max. $\Pcoops$}&\multicolumn{1}{p{2.1cm}|}{\centering Points with \\ min. $\Pcoops$} &\raisebox{-1.5ex}[0pt]{$\lDbi$ \text{(Mbps)}}&\raisebox{-1.5ex}[0pt]{$\uDbi$\text{(Mbps)}}\\ \hline
2\; &  $ \text b_1$, $ \text b_2$&\multicolumn{1}{p{2.1cm}|}{\centering e\textsubscript{1}, e\textsubscript{2}, e\textsubscript{3}, e\textsubscript{4} }&$\mathbb{G}(48.2,67.1)\times 3.67$ & $\mathbb{G}(48.2,r_k-48.2)\times 3.67$ \\ \hline
3\;&  \text o&$ \text d_1$, $ \text d_2$& $\mathbb{G}(67.1,67.1)\times 2.75 $ & $\mathbb{G}\big(\frac{r_k}{2},\frac{r_k}{2}\big)\times 2.75 $ \\ \hline
4\;& $ \text c_1$, $ \text c_2$&\multicolumn{1}{p{2cm}|}{\centering f\textsubscript{1}, f\textsubscript{2}, f\textsubscript{3}, f\textsubscript{4}}& $\mathbb{G}(48.2,74.7)\times 1.69 $& $\mathbb{G}(67.1,r_k-67.1)\times 1.69 $\\ \hline
5\;& \multicolumn{1}{p{2.3cm}|}{\centering e\textsubscript{1}, e\textsubscript{2}, e\textsubscript{3}, e\textsubscript{4}}&\multicolumn{1}{p{2cm}|}{\centering i\textsubscript{1}, i\textsubscript{2}, i\textsubscript{3}, i\textsubscript{4}}& $\mathbb{G}(67.1,74.7)\times 1.47$ &$\mathbb{G}(48.2,67.1)\times 1.47$\\ \hline
\end{tabular}
\label{T5}
\end{table*}
Averaging over the distribution of $r_k$ we obtain
\begin{subequations} \label{TD1}
\begin{gather}
    \mathcal{\overline L}^{\mathcal{D}_1}\le\mathcal{\overline T}^{\mathcal{D}_1}\le\mathcal{\overline U}^{\mathcal{D}_1}\\
    \shortintertext{where}
	\mathcal{\overline L}^{\mathcal{D}_1}=\int_{74.7}^{96.4} 	\mathcal{L}^{\mathcal{D}_1}(r) \frk \mathrm{d}r  \\
	\mathcal{\overline U}^{\mathcal{D}_1}=\int_{74.7}^{96.4} 	\mathcal{U}^{\mathcal{D}_1}(r) \frk \mathrm{d}r
\end{gather}
\end{subequations}
and $\mathcal{\overline T}^{\mathcal{D}_1}$ is the average throughput of a Type $\mathcal{D}$ link for the case where $74.7< r_k\le 96.4$.
\subsection{Bounds on the Cooperative Throughput for $96.4 < r_k \le 100$} \label{SubD2}
The analysis in this subsection is similar to what presented in Subsection \ref{SubD1} except that in this case a Tier $1$ helper cannot be used as mentioned in
Subsection \ref{sysmodD}.  Therefore, one has
\begin{equation}
\lDbi\le \mathcal{T}_\text{Coop}^{\mathcal{D}_2,i} \le \uDbi,\quad i=2,\ldots,5
\end{equation}
where $\lDbi$ and $\uDbi$, $i=2,\ldots,5$ are shown in Table \ref{T5} for each helper's tier.  Also shown in this table are the points at which the maximum and minimum $\Pcoops$ are achieved for each tier.
\begin{table*}[!t]
\caption{The simulation parameters.}
\small
\centering
    \begin{tabular}{ | c | c | c | c | c | c | c | c | c | c | c |}
    \hline
    Parameter & $P_\text{t}$ & $P_\text{th}$ &$ \alpha$& $\sigma_{\psi}$& $K$& RTS& CoopRTS& CTS& HTS& data   \\ \hline
    Value & $1$ mW &$-98$ dBm& $3$& $6$ dB& $-40$ dB& $352$ bits&$352$ bits& $304$ bits& $304$ bits& $1000$ bytes \\ \hline
    \end{tabular}
\label{T6}
\end{table*}
Hence, we can write
\begin{subequations} \label{boundsD2}
\begin{gather}
    \mathcal{L}^{\mathcal{D}_2}(r_k)\le\mathcal{T}^{\mathcal{D}_2}(r_k)\le\mathcal{U}^{\mathcal{D}_2}(r_k)\\
    \shortintertext{where}
	\mathcal{L}^{\mathcal{D}_2}(r_k)=\sum_{i=2}^5\mathcal{P}_{\mathcal{D},i}\,\mathcal{L}_\text{Coop}^{\mathcal{D}_2,i}+\bigg(1-\sum_{i=2}^5\mathcal{P}_{\mathcal{D},i}\bigg)\Ps(r_k)\times 1~\text{(Mbps)} \\
	\mathcal{U}^{\mathcal{D}_2}(r_k)=\sum_{i=2}^5\mathcal{P}_{\mathcal{D},i}\,\mathcal{U}_\text{Coop}^{\mathcal{D}_2,i}+\bigg(1-\sum_{i=2}^5\mathcal{P}_{\mathcal{D},i}\bigg)\Ps(r_k)\times 1~\text{(Mbps)}.
\end{gather}
\end{subequations}
Using \eqref{boundsD2}, we can find the upper and lower bounds of the average throughput of a Type $\mathcal{D}$ link for the case where $96.4<r_k\le100$ as
\begin{subequations}\label{TD2}
\begin{gather}
    \mathcal{\overline L}^{\mathcal{D}_2}\le\mathcal{\overline T}^{\mathcal{D}_2}\le\mathcal{\overline U}^{\mathcal{D}_2}\\
    \shortintertext{where}
	\mathcal{\overline L}^{\mathcal{D}_2}=\int_{96.4}^{100} 	\mathcal{L}^{\mathcal{D}_2}(r) \frk \mathrm{d}r  \\
	\mathcal{\overline U}^{\mathcal{D}_2}=\int_{96.4}^{100} 	\mathcal{U}^{\mathcal{D}_2}(r) \frk \mathrm{d}r.
\end{gather}
\end{subequations}

In summary, we can combine eqs. \eqref{throughputA}, \eqref{throughputB}, \eqref{TC}, \eqref{TD1} and \eqref{TD2} to bound the average throughput of our proposed CoopMAC protocol in the presence of shadowing as
\begin{subequations}\label{TD3}
\begin{gather}
    \mathcal{\overline L}\le\mathcal{\overline T}\le\mathcal{\overline U}\\
    \shortintertext{where}
	\mathcal{\overline L}=\mathcal{\overline L}^{\mathcal{D}_2}+\mathcal{\overline L}^{\mathcal{D}_1}+\mathcal{\overline L}^{\mathcal{C}}+\mathcal{T}^\mathcal{A}+\mathcal{T}^\mathcal{B}  \\
	\mathcal{\overline U}=\mathcal{\overline U}^{\mathcal{D}_2}+\mathcal{\overline U}^{\mathcal{D}_1}+\mathcal{\overline U}^{\mathcal{C}}+\mathcal{T}^\mathcal{A}+\mathcal{T}^\mathcal{B}.
\end{gather}
\end{subequations}
\begin{figure}[!t]
\centering
\includegraphics[height=85mm]{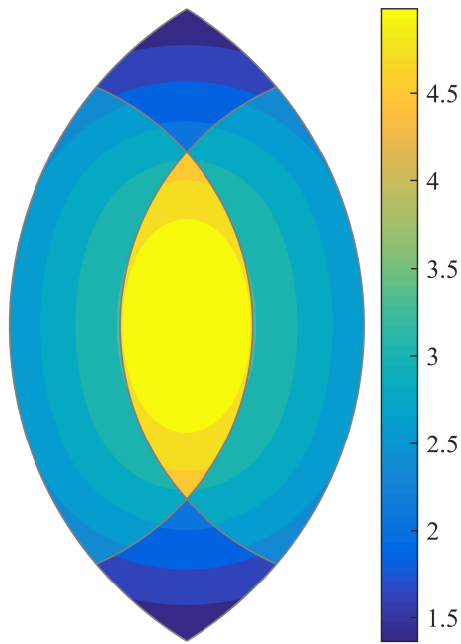}
\caption{The contour plot of the cooperative throughput of a Type $\mathcal{C}$ link achieved through Tier 1, 2 and 3 helpers for $67.1<r_k\le74.7$. The throughputs are in Mbps.}
\label{contc}
\end{figure}
\section{Numerical Results} \label{simulation}
We have used computer simulation to evaluate the throughput performance of our proposed CoopMAC scheme and illustrate its superiority over the conventional CoopMAC protocol in which the helpers are selected randomly (i.e., the locations of helpers are not taken into account).  As mentioned earlier, only Type $\mathcal{C}$ and Type $\mathcal{D}$ links can take advantage of cooperation and, therefore, we have only considered these link types in our analysis.  Table \ref{T6} shows the parameters that have been used in our computer simulations.  Throughout this section, we assume these parameters remain unchanged unless otherwise specified. The simulation results have been obtained using the Monte-Carlo method for two million independent realizations of a network whose nodes are distributed according to a two-dimensional PPP with density $0.0005\le\lambda\le 0.005$.
\par Fig. \ref{contc} shows a contour plot of the cooperative throughput achieved by a Type $\mathcal{C}$ link.  Clearly, the achievable throughput for most of Tier 1 helpers is greater than 4.5 Mbps (and, actually, very close to $\mathbb{G}\big(\frac{r_k}{2},\frac{r_k}{2}\big)\times5.5$ Mbps), and for a small fraction of these helpers the throughput is smaller than $4.5$ Mbps.
For Tier 2 helpers, the maximum achievable throughput is approximately $1$ Mbps less than the minimum throughput that can be achieved by a Tier 1 helper. This explains why a Tier 1 helper is superior to a Tier 2 helper in our proposed scheme. Note that the throughput achieved by Tier 3 helpers is less than $50\%$ of that of the Tier 1 helpers. For this reason, we give them the lowest priority in our proposed scheme.
%
\begin{figure}[!t]
\centering
\begin{tikzpicture}[scale=1.75,font=\tiny]
\begin{axis}[
    xlabel={ {\tiny Density of Nodes ($\lambda$)}},
    ylabel={{\tiny Average Cooperative Throughput (Mbps)}},
    xlabel style={yshift=.5em},
    ylabel style={yshift=-1.25em},
    xmin=0.0005, xmax=0.005,
    ymin=2.5, ymax=5.5,
    xtick={0.0005,0.001,0.0015,0.002,0.0025,0.003,0.0035,0.004,0.0045,0.005},
    ytick={2.5,3,3.5,4,4.5,5,5.5},
    yticklabel style = {font=\tiny,xshift=0.25ex},
  xticklabel style = {font=\tiny,yshift=0.25ex},
    legend pos=north west,
legend style = { at = {(0.5,0.75)}}
]

\addplot[
    color=blue,
    mark=none,
    ]
    coordinates {
    (0.0005,4.4403)(0.001,4.6002)(0.0015,4.7303)(0.002,4.8208)(0.0025,4.9003)(0.003,4.9506)(0.0035,5.0008)(0.004,5.0205)(0.0045,5.0408)(0.005,5.0556)
    };
\addplot[
    color=blue,
    mark=halfcircle,
    ]
    coordinates {
    (0.0005,4.3301)(0.001,4.4808)(0.0015,4.6302)(0.002,4.7404)(0.0025,4.8309)(0.003,4.9004)(0.0035,4.9505)(0.004,4.9806)(0.0045,5.0002)(0.005,5.0302)
    };
\addplot[
    color=blue,
       mark=star,
    ]
    coordinates {
    (0.0005,2.8505)(0.001,3.1008)(0.0015,3.2509)(0.002,3.3606)(0.0025,3.4306)(0.003,3.5008)(0.0035,3.5401)(0.004,3.5909)(0.0045,3.6200)(0.005,3.6403)
    };
\addplot[
    color=blue,
    mark=square,
    ]
    coordinates {
    (0.0005,2.7001)(0.001,2.8803)(0.0015,3.0003)(0.002,3.1005)(0.0025,3.1705)(0.003,3.2308)(0.0035,3.2801)(0.004,3.3101)(0.0045,3.3403)(0.005,3.3605)
    };
    \legend{{Upper Bound},{Proposed CoopMAC},{Conventional CoopMAC},{Lower Bound}}

\end{axis}
\end{tikzpicture}
\caption{The average throughput as a function of $\lambda$ for a Type $\mathcal{C}$ link making use of the proposed and conventional CoopMAC
protocols.}
\label{figc}
\end{figure}
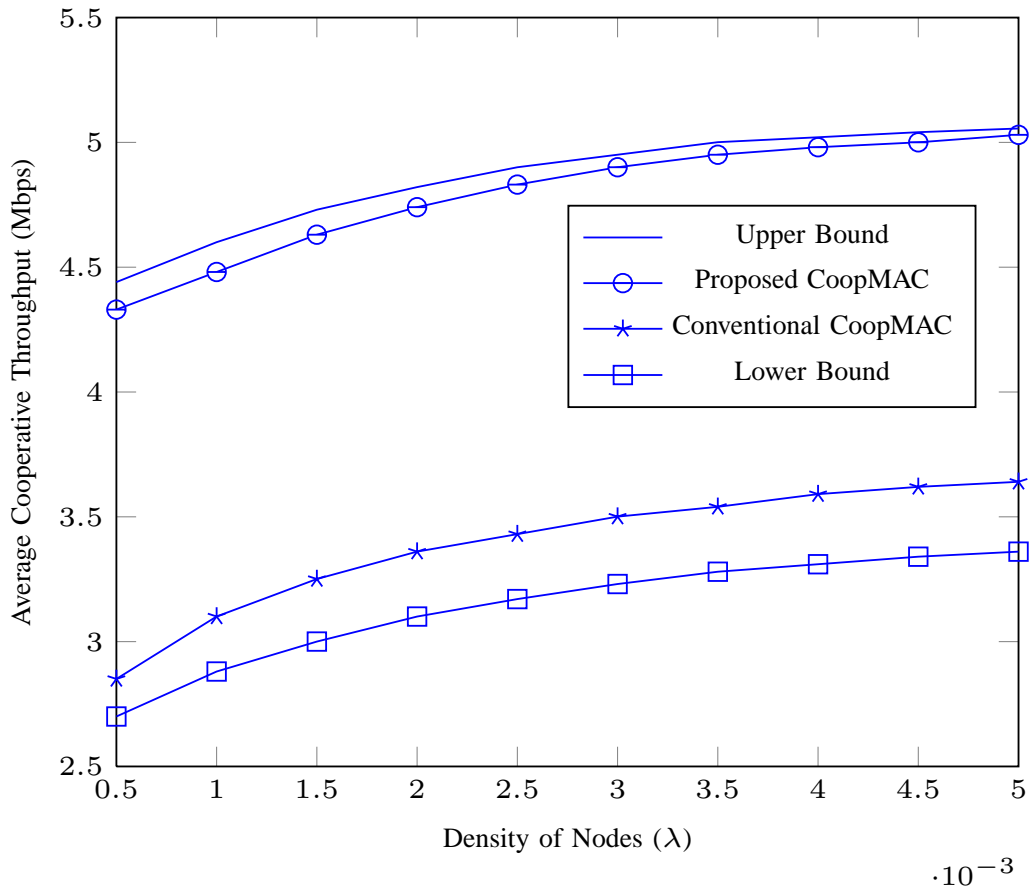
\par The average throughput of a Type $\mathcal{C}$ link making use of the proposed and the conventional CoopMAC schemes as a function of $\lambda$ are shown in Fig. \ref{figc}.  Also shown in this figure are the upper and lower bounds derived in \eqref{TC} for a Type $\mathcal{C}$ link.  As seen in this figure, the throughput improvement due to proposed scheme is quite significant. Observe that when $\lambda$ increases, the throughput of the proposed scheme becomes very close to the upper bound.  This is because when $\lambda$ increases the chance of finding a helper near the best helper (i.e., a helper located halfway between source and destination) also increases.  Fig. \ref{figc} also shows that for a Type $\mathcal{C}$ link utilizing the conventional CoopMAC protocol the average throughput is slightly larger than the lower bound.  To explain this, we note from Fig. \ref{contc} that a randomly selected helper is more likely to be from Tiers 2 and 3 and the cooperative throughput that can be achieved through a helper from these tiers is generally closer to the lower bound than the upper bound.
\begin{figure}[!t]
\centering
\begin{subfigure}[t]{.5\linewidth}
  \centering
  \includegraphics[height=85mm,clip]{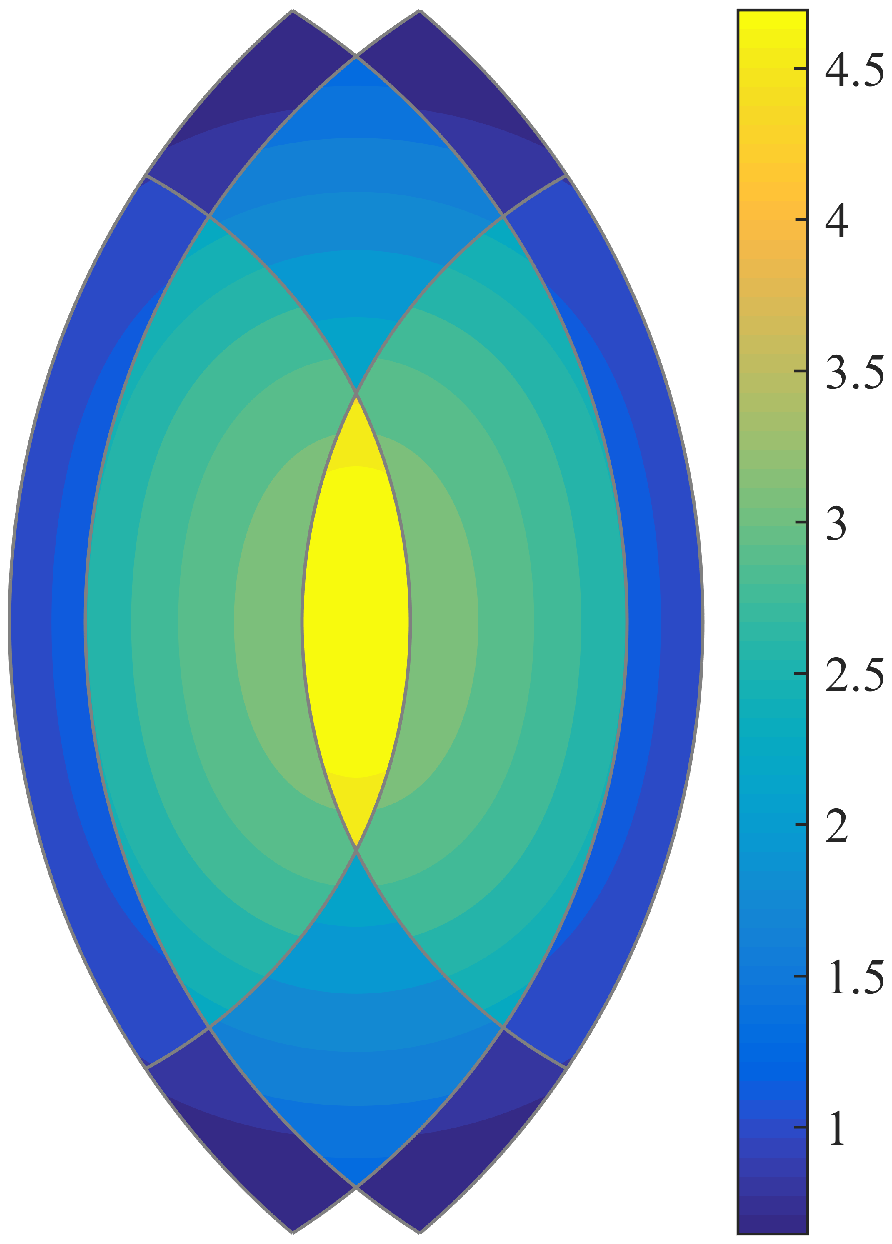}
  \caption{}
  \label{contda}
\end{subfigure}~~%
\begin{subfigure}[t]{.5\linewidth}
  \centering
  \includegraphics[height=85mm,clip]{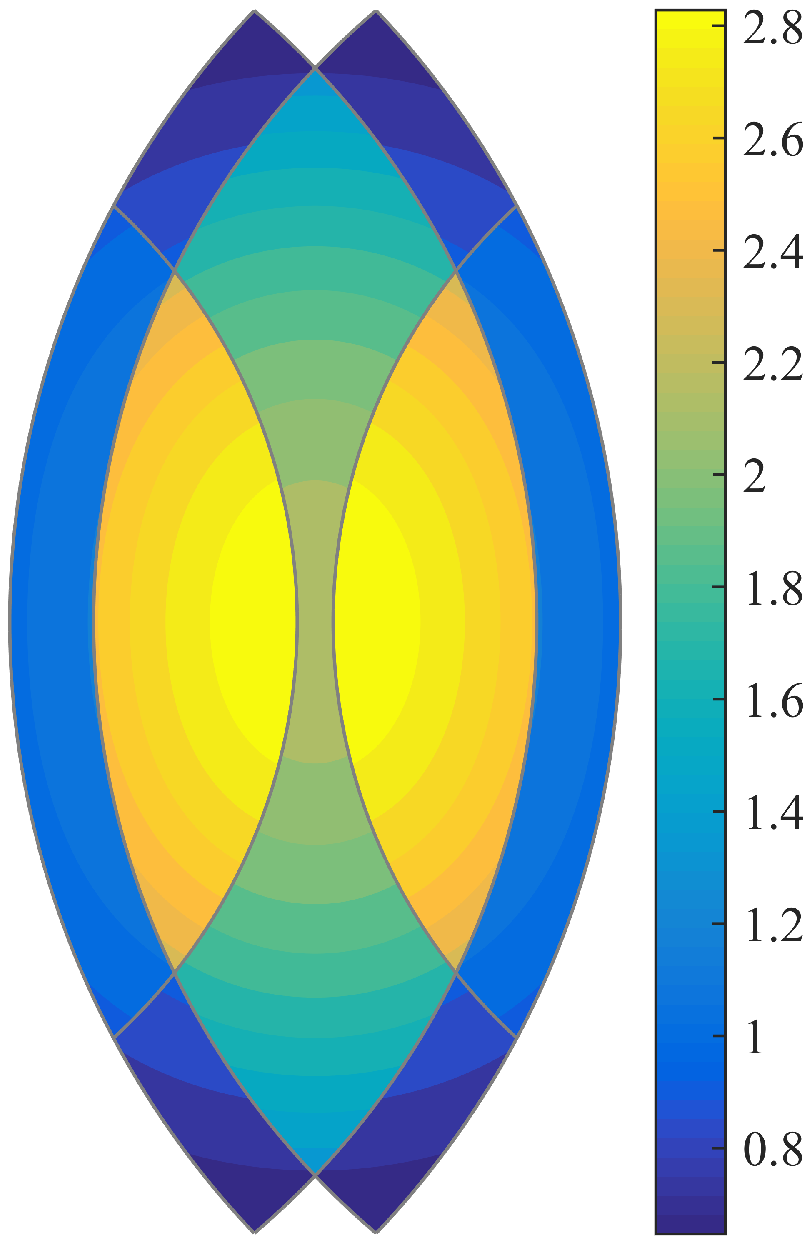}
  \caption{}
  \label{contdb}
\end{subfigure}
\caption{The contour plots of the cooperative throughput of a Type $\mathcal{D}$ link for (a)~$74.7<r_k\le96.4$, and (b)~$96.4<r_k\le100$. The throughputs are in Mbps.}
\label{contd}
\end{figure}
\par The cooperative throughputs of a Type $\mathcal{D}$ link are illustrated as contour maps in Fig. \ref{contda} for $74.7<r_k\le96.4$ and in Fig. \ref{contdb} and for $96.4<r_k\le100$.  Clearly, for the case where $74.7< r_k\le 96.4$, the achievable cooperative throughput can be as large as $4.6$ Mbps whereas for $96.4<r_k\le100$ the maximum cooperative throughput is approximately $2.8$ Mbps.  As mentioned in Section \ref{SubD1}, this difference has its roots in the fact that in the former case a Tier 1 helper can be taken advantage of whereas in the latter it cannot.  Note, importantly, that in Fig. \ref{contda} the maximum cooperative throughput can be achieved through the helpers that are no farther than $48.2$ meters from the source and destination nodes.  Since only a small number of helpers have this property, a selection scheme that does not account for the locations of helpers is unlikely to select one of these helpers and, thus, achieve the maximum throughput.

%
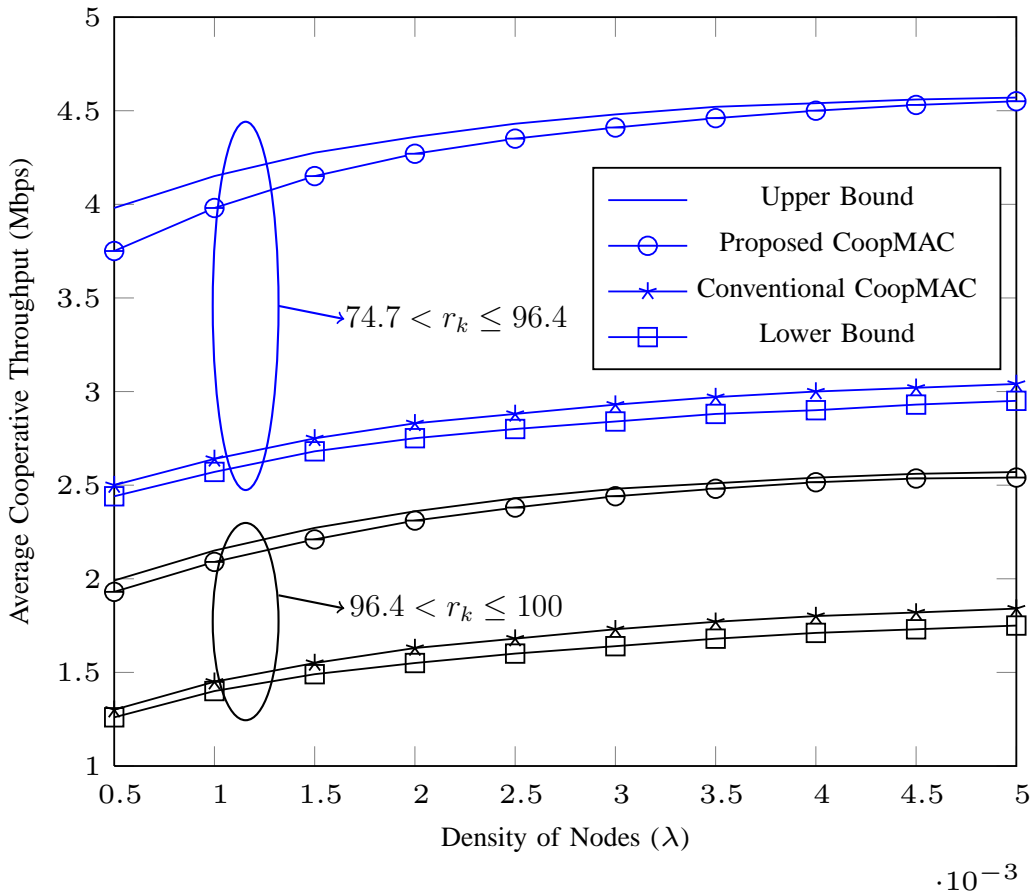
\begin{figure}[!t]
\centering
\begin{tikzpicture}[scale=1.75]
\begin{axis}[
    xlabel={ {\tiny Density of Nodes ($\lambda$)} },
    ylabel={{\tiny Average Cooperative Throughput (Mbps)}},
     xlabel style={yshift=.5em},
        ylabel style={yshift=-1.25em},
    xmin=0.0005, xmax=0.005,
    ymin=1, ymax=5,
    xtick={0.0005,0.001,0.0015,0.002,0.0025,0.003,0.0035,0.004,0.0045,0.005},
    ytick={1,1.5,2,2.5,3,3.5,4,4.5,5,5.5},
    yticklabel style = {font=\tiny,xshift=0.25ex},
 xticklabel style = {font=\tiny,xshift=0.25ex},
    legend pos=north west,
legend style = { at = {(0.53,0.8)}}
]
\addplot[
    color=blue,
    mark=none,
    ]
    coordinates {
    (0.0005,3.9807)(0.001,4.1510)(0.0015,4.2759)(0.002,4.3603)(0.0025,4.4306)(0.003,4.4802)(0.0035,4.5209)(0.004,4.5401)(0.0045,4.5600)(0.005,4.5703)
    };
\addplot[
    color=blue,
    mark=halfcircle,
    ]
    coordinates {
    (0.0005,3.7507)(0.001,3.9805)(0.0015,4.1508)(0.002,4.2702)(0.0025,4.3509)(0.003,4.4102)(0.0035,4.4603)(0.004,4.5001)(0.0045,4.5308)(0.005,4.5505)
    };
\addplot[
    color=blue,
       mark=star,
    ]
    coordinates {
    (0.0005,2.5003)(0.001,2.6402)(0.0015,2.7500)(0.002,2.8304)(0.0025,2.8805)(0.003,2.9302)(0.0035,2.9705)(0.004,3.0002)(0.0045,3.0202)(0.005,3.0406)
    };
\addplot[
    color=blue,
    mark=square,
    ]
    coordinates {
    (0.0005,2.4407)(0.001,2.5708)(0.0015,2.6808)(0.002,2.7509)(0.0025,2.8001)(0.003,2.8401)(0.0035,2.8805)(0.004,2.9004)(0.0045,2.9304)(0.005,2.9505)
    };
    \legend{{\tiny Upper Bound},{\tiny Proposed CoopMAC},{\tiny Conventional CoopMAC},{\tiny Lower Bound}}
 \addplot[
    color=black,
    mark=none,
    ]
    coordinates {
    (0.0005,1.9909)(0.001,2.1510)(0.0015,2.2718)(0.002,2.3606)(0.0025,2.4301)(0.003,2.4809)(0.0035,2.5104)(0.004,2.5401)(0.0045,2.5607)(0.005,2.5702)
    };
\addplot[
    color=black,
    mark=halfcircle,
    ]
    coordinates {
    (0.0005,1.9302)(0.001,2.0901)(0.0015,2.2103)(0.002,2.3109)(0.0025,2.3804)(0.003,2.4409)(0.0035,2.4808)(0.004,2.5156)(0.0045,2.5356)(0.005,2.5407)
    };
\addplot[
    color=black,
       mark=star,
    ]
    coordinates {
    (0.0005,1.3005)(0.001,1.4501)(0.0015,1.5501)(0.002,1.6301)(0.0025,1.6804)(0.003,1.7304)(0.0035,1.7704)(0.004,1.8009)(0.0045,1.8200)(0.005,1.8407)
    };
\addplot[
    color=black,
    mark=square,
    ]
    coordinates {
    (0.0005,1.2601)(0.001,1.4004)(0.0015,1.4904)(0.002,1.5501)(0.0025,1.6005)(0.003,1.6400)(0.0035,1.6801)(0.004,1.7110)(0.0045,1.7302)(0.005,1.7502)
    };
\end{axis}
\draw [thick](1,1.1) ellipse (0.25cm and 0.75cm);
\draw [color=blue,thick] (1,3.5) ellipse (0.25cm and 1.4cm);
\draw [->,thick](1.25,1.3)--(1.75,1.2);
\node [draw=none] (v1) at (2.6,1.2) {$96.4<r_k\le 100$};
\draw [->,thick,color=blue](1.25,3.5)--(1.75,3.4);
\node [draw=none] (v1) at (2.6,3.4) {$74.7<r_k\le 96.4$};
\end{tikzpicture}
\caption{The average cooperative throughput as a function of $\lambda$ for a Type $\mathcal{D}$ link making use of the proposed and conventional CoopMAC
protocols.}
\label{figdf}
\end{figure}
\par The average cooperative throughput of a Type $\mathcal{D}$ link making use of the proposed and conventional CoopMAC protocol is shown in Fig. \ref{figdf}.  Both $74.7<r_k\le96.4$ and $96.4<r_k\le100$ cases are considered.  Similar to Fig. \ref{figc}, in this figure the average throughput of the proposed scheme is close to the upper bound particularly when $\lambda$ approaches $0.005$.  Note that the average throughput of conventional CoopMAC scheme is slightly larger than the lower bound.  This is due to the fact that most of the cooperative throughputs illustrated in Figs. \ref{contda} and \ref{contdb} are closer to the lower bound than the upper bound.  Hence, random selection of the helpers results in an average throughput that is close to the lower bound.

\par
\begin{figure}[!t]
\centering
\begin{tikzpicture}[scale=1.75,font=\tiny]
\begin{axis}[
    xlabel={ {\tiny Density of Nodes ($\lambda$)} },
    ylabel={{\tiny Average  Throughput (Mbps)}},
     xlabel style={yshift=.5em},
     ylabel style={yshift=-1.25em},
    xmin=0.0005, xmax=0.005,
    ymin=5, ymax=9.5,
    xtick={0.0005,0.001,0.0015,0.002,0.0025,0.003,0.0035,0.004,0.0045,0.005},
    ytick={5,5.5,6,6.5,7,7.5,8,8.5,9,9.5},
    yticklabel style = {font=\tiny,xshift=0.25ex},
    xticklabel style = {font=\tiny,yshift=0.25ex},
    legend pos=north west,
legend style = { at = {(0.5,0.35)}}
]

\addplot[
    color=blue,
    mark=none,
    ]
    coordinates {
    (0.0005,7.0502)(0.001,7.5500)(0.0015,7.9503)(0.002,8.3004)(0.0025,8.5605)(0.003,8.7808)(0.0035,8.9402)(0.004,9.0503)(0.0045,9.1303)(0.005,9.1810)
    };
\addplot[
    color=blue,
    mark=halfcircle,
    ]
    coordinates {
    (0.0005,6.7754)(0.001,7.3254)(0.0015,7.8054)(0.002,8.1954)(0.0025,8.4954)(0.003,8.7254)(0.0035,8.8954)(0.004,9.0154)(0.0045,9.0904)(0.005,9.1404)
    };
\addplot[
    color=blue,
       mark=star,
    ]
    coordinates {
    (0.0005,5.4007)(0.001,6.0007)(0.0015,6.4001)(0.002,6.7510)(0.0025,7.0008)(0.003,7.2304)(0.0035,7.4103)(0.004,7.5301)(0.0045,7.6006)(0.005,7.6405)
    };
\addplot[
    color=blue,
    mark=square,
    ]
    coordinates {
     (0.0005,5.2004)(0.001,5.7500)(0.0015,6.1508)(0.002,6.4800)(0.0025,6.7308)(0.003,6.9506)(0.0035,7.1007)(0.004,7.2305)(0.0045,7.3002)(0.005,7.3507)
    };
    \legend{{\tiny Upper Bound},{\tiny Proposed CoopMAC},{\tiny Conventional CoopMAC},{\tiny Lower Bound}}

\end{axis}
\end{tikzpicture}
\caption{The average throughputs of all link types as a function of $\lambda$ for the proposed and conventional CoopMAC
schemes.}
\label{figtf}
\end{figure}
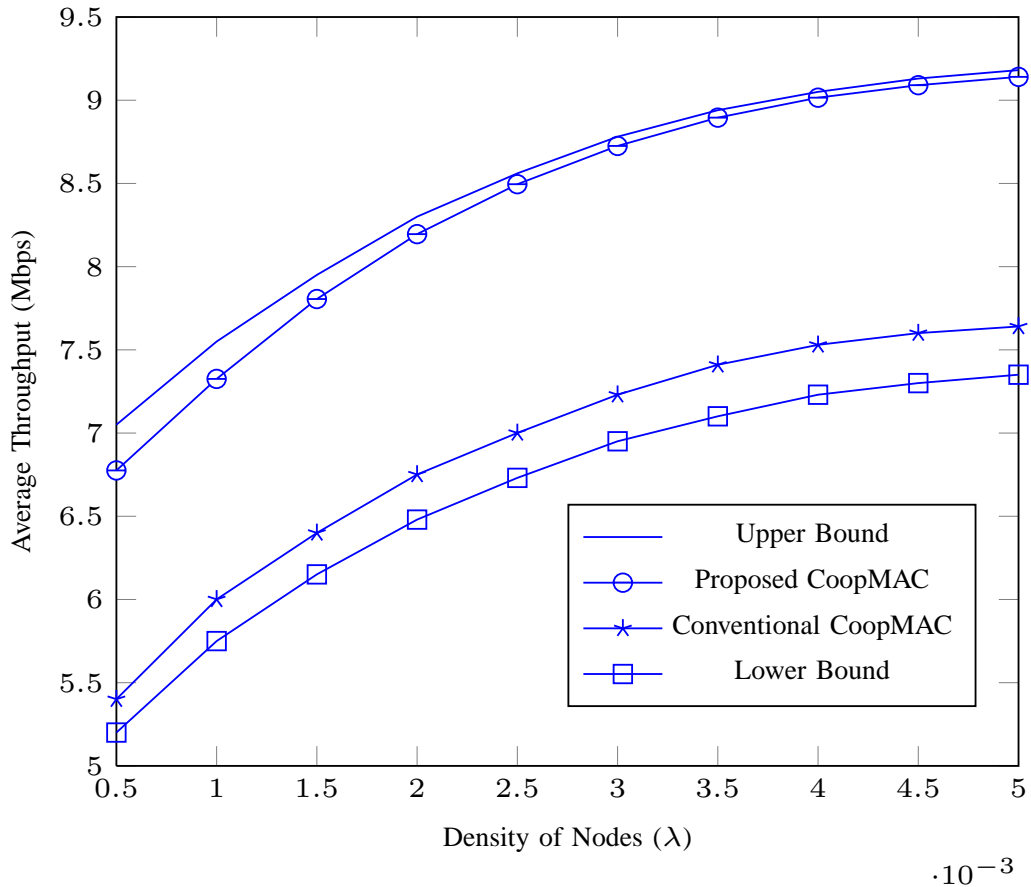

 Fig. \ref{figtf} shows the average throughput as a function of $\lambda$ achieved by the proposed and conventional CoopMAC schemes for all link types.  Observe that in this case, Type $\mathcal{A}$ and $\mathcal{B}$ links also contribute to the average throughput.  Therefore, the average throughput is approximately twice as large as it was for Type $\mathcal{C}$ and $\mathcal{D}$ links.  Similar to Figs. \ref{figc} and \ref{figdf}, by increasing the density of nodes, the average throughput of the proposed scheme becomes closer to the upper bound in contrast to the conventional CoopMAC whose throughput performance does not improve much.
\section{Conclusion}\label{conclude}
In this paper, we considered a CoopMAC network based on the IEEE 802.11b Standard, and studied its throughput performance in the presence of shadowing and spatially distributed random nodes.  We first identified four link types according to their achievable throughput and divided the potential helpers for each link type into several tiers based on the cooperative throughput that they can provide.  Then, the locus of the helpers in each tier were determined using simple algebraic expressions.  In our proposed CoopMAC protocol, the helpers with the lowest tier index have the highest priority to be selected for cooperation. We derived upper and lower bounds on the average throughput of different link types in the network.  Our numerical results illustrated the superiority of our scheme over the conventional CoopMAC protocol.  Indeed, in all the examined scenarios the average throughput of the proposed scheme was very close to the upper bound while the average throughput of the conventional scheme was slightly larger than the lower bound.



\bibliographystyle{IEEEtran}
\bibliography{Nikbakht}

\begin{thebibliography}{10}
\providecommand{\url}[1]{#1}
\csname url@samestyle\endcsname
\providecommand{\newblock}{\relax}
\providecommand{\bibinfo}[2]{#2}
\providecommand{\BIBentrySTDinterwordspacing}{\spaceskip=0pt\relax}
\providecommand{\BIBentryALTinterwordstretchfactor}{4}
\providecommand{\BIBentryALTinterwordspacing}{\spaceskip=\fontdimen2\font plus
\BIBentryALTinterwordstretchfactor\fontdimen3\font minus
  \fontdimen4\font\relax}
\providecommand{\BIBforeignlanguage}[2]{{%
\expandafter\ifx\csname l@#1\endcsname\relax
\typeout{** WARNING: IEEEtran.bst: No hyphenation pattern has been}%
\typeout{** loaded for the language `#1'. Using the pattern for}%
\typeout{** the default language instead.}%
\else
\language=\csname l@#1\endcsname
\fi
#2}}
\providecommand{\BIBdecl}{\relax}
\BIBdecl

\bibitem{IEEE802.11b}
``{IEEE} standard for information technology--telecommunications and
  information exchange between systems--local and metropolitan area
  networks--specific requirements part 11: wireless {LAN} medium access control
  ({MAC}) and physical layer ({PHY}) specifications,'' \emph{{IEEE Std}
  802.11}, pp. 1--528, 1999.

\bibitem{Korakis}
P.~Liu, Z.~Tao, S.~Narayanan, T.~Korakis, and S.~S. Panwar, ``{CoopMAC}:a
  cooperative {MAC} for wireless {LAN}s,'' \emph{{IEEE} Trans. Commun.},
  vol.~25, no.~2, pp. 340--354, Feb. 2007.

\bibitem{CODE}
K.~Tan, Z.~Wan, H.~Zhu, and J.~Andrian, ``{CODE}: Cooperative medium access for
  multirate wireless ad hoc network,'' in \emph{{IEEE} Sensor Mesh Ad Hoc
  Commun. Networks Conf. (SECON)}, San Diego, CA, Jun. 2007, pp. 1--10.

\bibitem{forward}
S.~Narayanan and S.~S. Panwar, ``To forward or not to forward--that is the
  question,'' \emph{{Springer} Wireless Personal Commun.}, vol.~43, no.~1, pp.
  65--87, Feb. 2007.

\bibitem{CoopRTS/CTS}
X.~He and F.~Y. Li, ``{Cooperative} {RTS/CTS} {MAC} with relay selection in
  distributed wireless networks,'' in \emph{{IEEE} Ultra Modern Telecommun.
  Workshops Conf. (ICUMT)}, St. Petersburg, Oct. 2009, pp. 1--8.

\bibitem{WCNC}
P.~Ju, W.~Song, and D.~Zhou, ``An enhanced cooperative {MAC} protocol based on
  perceptron trainings,'' in \emph{{IEEE} Wireless Commun. Networking Conf.},
  Shanghai, China, Apr. 2013, pp. 404--409.

\bibitem{PRCSMA}
J.~Alonso-Zrate, E.~Kartsakli, C.~Verikoukis, and L.~Alonso, ``Persistent
  {RCSMA}: A {MAC} protocol for a distributed cooperative {ARQ} scheme in
  wireless networks,'' \emph{{IEEE} EURASIP J. Advances Signal Process.}, vol.
  2008, pp. 1--13, Apr. 2008.

\bibitem{distributed}
C.~Chou, J.~Yang, and D.~Wang, ``{Cooperative} {MAC} protocol with automatic
  relay selection in distributed wireless networks,'' in \emph{{IEEE} Pervasive
  Computing Commun. Workshops Conf. (PerCom)}, White Plains, NY, Mar. 2007, pp.
  526--531.

\bibitem{relayselection}
T.~Jamal and P.~Mendes, ``Relay selection approaches for wireless cooperative
  networks,'' in \emph{{IEEE} Wireless Mobile Computing Networking Commun.
  Conf. (WiMob)}, Niagara Falls, ON, Oct. 2010, pp. 661--668.

\bibitem{Wang15}
X.~Wang and J.~Li, ``Improving the network lifetime of {MANETs} through
  cooperative {MAC} protocol design,'' \emph{IEEE Trans. Parallel Dist.
  Systems}, vol.~26, no.~4, pp. 1010--1020, Apr. 2015.

\bibitem{Ju15}
P.~Ju and W.~Song, ``Repeated game analysis for cooperative {MAC} with
  incentive design for wireless networks,'' \emph{{IEEE} Trans. Veh. Technol.},
  vol.~65, no.~7, pp. 5045--5059, Jul. 2016.

\bibitem{shadow}
M.~D. Renzo, J.~Alonso-Zarate, L.~Alonso, and C.~Verikoukis, ``On the impact of
  shadowing on the performance of cooperative medium access control
  protocols,'' in \emph{{IEEE} Global Telecommun. Conf.}, Houston, TX, Dec.
  2011, pp. 1--6.

\bibitem{shadow2}
A.~Antonopoulos, M.~Renzo, and C.~Verikoukis, ``Effect of realistic channel
  conditions on the energy efficiency of network coding-aided cooperative {MAC}
  protocols,'' \emph{{IEEE} Trans. Wireless Commun.}, vol.~20, no.~5, pp.
  76--84, Oct. 2013.

\bibitem{shadownc}
A.~S. Lalos, M.~D. Renzo, L.~Alonso, and C.~Verikoukis, ``Impact of correlated
  log-normal shadowing on two-way network coded cooperative wireless
  networks,'' \emph{{IEEE} Commun. Lett.}, vol.~17, no.~9, pp. 1738--1741, Sep.
  2013.

\bibitem{Wang_2011}
H.~Wang, S.~Ma, and T.-S. Ng, ``On performance of cooperative communication
  systems with spatial random relays,'' \emph{{IEEE} Trans. Commun.}, vol.~59,
  no.~4, pp. 1190--1199, Apr. 2011.

\bibitem{behnad_TCOM_2013}
A.~Behnad, A.~M. Rabiei, and N.~Beaulieu, ``Performance analysis of
  opportunistic relaying in a {P}oisson field of amplify-and-forward relays,''
  \emph{{IEEE} Trans. Commun.}, vol.~61, no.~1, pp. 97--107, Jan. 2013.

\bibitem{Goldsmith}
A.~Goldsmith, \emph{Wireless Communications}.\hskip 1em plus 0.5em minus
  0.4em\relax New York: Cambridge University Press, 2005.

\bibitem{Haenggi2005}
M.~Haenggi, ``On distances in uniformly random networks,'' \emph{{IEEE} Trans.
  Inform. Theory}, vol.~51, no.~10, pp. 3584--3586, Oct. 2005.

\bibitem{Moller}
J.~{M{\o}ller} and R.~P. Waagepetersen, \emph{Statistical Inference and
  Simulation for Spatial Point Processes}.\hskip 1em plus 0.5em minus
  0.4em\relax Chapman \& Hall/CRC, 2004.

\bibitem{boyd}
S.~Boyd and L.~Vandenberghe, \emph{Convex Optimization}.\hskip 1em plus 0.5em
  minus 0.4em\relax Cambridge University Press, 2004.

\end{thebibliography}
\appendices
\section{Proof of Lemma \ref{lemma1}} \label{appendix1}
Recalling from Table \ref{T2} that the transmission rates of all Tier 1 helpers for a Type $\mathcal{C}$ link are equal to $5.5$ Mbps, and that  $\mathcal{T}_\text{Coop}^{\mathcal{C},1}=\Rcoop\PcoopsCa$, we only need to find the maximum and minimum of $\PcoopsCa$.
To this end, we first note from Fig. \ref{fig2} that a helper is in $\mathscr{U}_1$ if
\begin{subequations}
\begin{align} \label{cond1}
\dsh &\le 48.2 \\\label{cond2} \dhd &\le 48.2 \\\label{cond3} r_k &\le \dsh+\dhd.
\end{align}
\end{subequations}
In order to maximize $\PcoopsCa=\mathbb{G}(\dsh,\dhd)$, one has to minimize the arguments of both $\mathbb{Q}$--functions in \eqref{PSuccessDef} owing to the fact that the Gaussian $\mathbb{Q}$--function is strictly decreasing in its argument. Hence, one has to minimize $\dsh$ and $\dhd$ provided that the constraints \eqref{cond1} through \eqref{cond3} are satisfied.  We first use proof by contradiction to show that the helper which maximizes $\PcoopsCa$ should be located on SD line segment in Fig. \ref{fig2}. We then prove that the best helper is indeed located halfway between S and D.
\par Suppose that H\textsuperscript{*} is a helper in $\mathscr{U}_1$ that maximizes $\PcoopsCa$ and is not located on SD line segment, i.e., $r_k < d_\text{SH\textsuperscript{*}}+d_\text{H\textsuperscript{*}D}$.
Assume now that H$^\perp$ is the projection of H\textsuperscript{*} on SD line segment and, thus,  $r_k = d_{\text{SH}^\perp}+d_\text{H$^\perp$D}$.  It is clear that, $d_\text{SH$^\perp$}<d_\text{SH\textsuperscript{*}}\le 48.2$ and $d_\text{H$^\perp$D}<d_\text{H\textsuperscript{*}D}\le 48.2$. Recalling that $\mathbb{Q}(x)$ is a monotonically decreasing function of $x$, we see from \eqref{PSuccessDef} that $\mathbb{G}(d_\text{SH\textsuperscript{*}},d_\text{H\textsuperscript{*}D})<\mathbb{G}(d_\text{SH$^\perp$},d_\text{H$^\perp$D})$.
Hence, our initial assumption that H\textsuperscript{*} maximizes $\PcoopsCa$ is wrong and the helper with maximum $\PcoopsCa$ has to be located on SD line segment, i.e., \eqref{cond3} should be changed to $\dsh+\dhd=r_k$ for this helper.
Substituting for $\dhd$ by $r_k-\dsh$ in \eqref{PSuccessDef} we obtain
\begin{equation} \label{Psucc2}
\mathbb{G}(\dsh,r_k-\dsh)=  \mathbb{Q}\left(\nu+\mu \log_{10}(\dsh)\right) \times \mathbb{Q}\left(\nu+\mu \log_{10}(r_k-\dsh)\right).
\end{equation}
Observe that for $0<\dsh<r_k$, $\mathbb{Q}\left(\nu+\mu \log_{10}(\dsh)\right)$ is a decreasing function of $\dsh$ whereas $\mathbb{Q}\left(\nu+\mu \log_{10}(r_k-\dsh)\right)$ is an increasing function of $\dsh$.  In addition, the arguments of both $\mathbb{Q}$--functions are negative and, thus, they are both concave functions of $\dsh$.  Consequently, the product of the $\mathbb{Q}$--functions in \eqref{Psucc2} is a concave function of $\dsh$ provided that $0<\dsh<r_k$ \cite[Exercise 3.32 (b)]{boyd}. Differentiating the right of \eqref{Psucc2} with respect to $\dsh$ and equating it to zero we obtain $\dsh=\frac{r_k}{2}$.  Thus, the maximum cooperative throughput in this case is obtained when the helper is located halfway between S and D (point M in Fig \ref{fig2}). Note that this result is optimum because it maximizes $\PcoopsCa$ and satisfies \eqref{cond1} through \eqref{cond3}.
\par To obtain the minimum value of $\PcoopsCa$, one should maximize the arguments of both $\mathbb{Q}$--functions in \eqref{PSuccessDef} so that the inequality constraints given in \eqref{cond1} to \eqref{cond3} are satisfied.  Considering the fact that $67.1\le r_k\le 74.7$, this occurs when $\dsh=\dhd=48.2$ m, i.e., the helper is located on  K\textsubscript{1} or K\textsubscript{2} in Fig. \ref{fig2}.
In summary, we can write
\begin{equation}
\mathbb{G}(48.2,48.2) \le \mathcal{P}^{\text{Succ},\mathcal{C},1}_{\text{Coop}} \le  \mathbb{G}\Big(\frac{r_k}{2},\frac{r_k}{2}\Big)
\end{equation}
which results in \eqref{TC1Coop}.

\section{Proof of Lemma \ref{lemma2}} \label{appendix2}
As shown in Fig. \ref{fig3}, a Tier 2 helper should be located either in $\mathscr{U}_{2,1}$ which is characterized as
\begin{subequations} \label{V2,1}
\begin{align}
~\dsh &\le 48.2 \\
48.2\le \dhd &\le 67.1\\
r_k &\le \dsh+\dhd
\end{align}
\end{subequations}
or in $\mathscr{U}_{2,2}$ characterized as
\begin{subequations} \label{V2,2}
\begin{align}
~\dhd &\le 48.2 \\
48.2\le \dsh &\le 67.1\\
r_k &\le \dsh+\dhd.
\end{align}
\end{subequations}
Similar to the proof given for a Tier 1 helper, we first show that the maximum $\PcoopsCb$ is achieved through a helper that is located on the SD line segment.  Again, we use a proof by contradiction.  Assume that H\textsuperscript{*} is a helper in $\mathscr{U}_{2,1}$ through which the maximum $\PcoopsCb$ is achieved.  Also assume that H$'$ is another helper in $\mathscr{U}_{2,1}$ which is located on the SD line segment such that $d_\text{H\textsuperscript{*}D}=d_\text{H$'$D}$ and $\mathbb{G}(d_\text{SH\textsuperscript{*}},d_\text{H\textsuperscript{*}D})>\mathbb{G}(d_\text{SH$'$},d_\text{H$'$D})$.  It is clear that $r_k < d_\text{SH\textsuperscript{*}}+d_\text{H\textsuperscript{*}D}$ and $r_k = d_\text{SH$'$}+d_\text{H$'$D}$.  Recalling that $d_\text{H\textsuperscript{*}D}=d_\text{H$'$D}$, one can readily see that $d_\text{SH$'$}<d_\text{SH\textsuperscript{*}}$ or analogously
\begin{equation}
\mathbb{Q}\left(\nu+\mu \log_{10}( d_\text{SH\textsuperscript{*}})\right)< \mathbb{Q}\left(\nu+\mu \log_{10}(d_{\text{SH}'})\right)
\end{equation}
which follows from the fact that the Gaussian $\mathbb{Q}$--function is monotonically decreasing in its argument. In consequence, using \eqref{PSuccessDefb} we can obtain $\mathbb{G}(d_\text{SH\textsuperscript{*}},d_\text{H\textsuperscript{*}D})<\mathbb{G}(d_\text{SH$'$},d_\text{H$'$D})$ which contradicts our initial assumption that $\mathbb{G}(d_\text{SH\textsuperscript{*}},d_\text{H\textsuperscript{*}D})>\mathbb{G}(d_\text{SH$'$},d_\text{H$'$D})$. This argument is also true for the case where the helpers are located in $\mathscr{U}_{2,2}$.  As a result, the helper with maximum $\PcoopsCb$ must be located on the SD line segment.  Hence, the maximum  $\PcoopsCb$ for the best helper should be obtained from \eqref{Psucc2}.  As seen in Appendix \ref{appendix1}, the product of the $\mathbb{Q}$--functions on the right of \eqref{Psucc2} is a concave function of $\dsh$ whose maximum is attained at $\dsh=\frac{r_k}{2}$.  For a Tier 2 helper that is located on the SD line segment, $\dsh$ cannot be equal to $\frac{r_k}{2}$.   However, considering the concavity of $\mathbb{G}(\dsh,r_k-\dsh)$ in $\dsh$ for $0<\dsh<r_k$, we conclude that the maximum of $\mathbb{G}(\dsh,r_k-\dsh)$ is attained at $\dsh=r_k-48.2$ (M\textsubscript{1} in Fig. \ref{fig4}), or at $\dsh=48.2$ (M\textsubscript{2} in Fig. \ref{fig4}).
\par To obtain the minimum of $\PcoopsCb$ we note from \eqref{PSuccessDef} that the $\mathbb{Q}$--functions are both minimum when their arguments are maximum. This minimum is attained at $\dsh=48.2$ and $\dhd=67.1$ when the helper is located in $\mathscr{U}_{2,1}$ (K\textsubscript{1} and K\textsubscript{2} in Fig. \ref{fig4}), or at $\dsh=67.1$ and $\dhd=48.2$ when the helper is located in $\mathscr{U}_{2,2}$ (K\textsubscript{3} and K\textsubscript{4} in Fig. \ref{fig4}).
Thus, $\PcoopsCb$ can be bounded as
\begin{equation}
\mathbb{G}(48.2,67.1) \le \mathcal{P}^{\text{Succ},\mathcal{C},2}_{\text{Coop}} \\\le  \mathbb{G}(48.2,r_k-48.2)
\end{equation}
which leads to \eqref{TC2Coop}.

\end{document}